\documentclass[letterpaper,10pt]{article}
\usepackage[margin=1in]{geometry}
\usepackage{lmodern}
\usepackage[T1]{fontenc}
\usepackage[backend=biber,sorting=none,style=vancouver,maxcitenames=10]{biblatex}
\usepackage{wrapfig}
\usepackage{amsmath}
\usepackage{comment}
\usepackage[separate-uncertainty=true,multi-part-units=single,range-phrase=--,range-units=single,group-separator={,},group-minimum-digits=4]{siunitx}
\usepackage{caption}
\usepackage{subcaption}
\usepackage{array}
\usepackage{bm}
\usepackage{graphicx}
\usepackage{authblk}
\usepackage{float}
\usepackage{amssymb}
\usepackage{xspace}
\usepackage{xr}
\usepackage{filecontents}

\usepackage[bookmarks]{hyperref}

\DeclareFieldFormat{doi}{%
  \mkbibacro{DOI}\addcolon\space
  \ifhyperref
    {\href{http://dx.doi.org/#1}{\nolinkurl{#1}}}
    {\nolinkurl{#1}}}

\renewcommand{\vec}[1]{\bm{\mathrm{#1}}}

\makeatletter
\newcommand*{\addFileDependency}[1]{% argument=file name and extension
  \typeout{(#1)}
  \@addtofilelist{#1}
  \IfFileExists{#1}{}{\typeout{No file #1.}}
}
\makeatother

\newcommand{\ourtool}{CardioFit\xspace}

\begin{document}
% title for Interface must be <=150 chars; this is 130
\title{\ourtool: A WebGL-Based Tool for Fast and Efficient Parameterization of Cardiac Action Potential Models to Fit User-Provided Data}

\author[1,*]{Darby I.\ Cairns}
\author[1,*]{Maxfield R.\ Comstock}
\author[2]{Flavio H.\ Fenton}
\author[1]{Elizabeth M.\ Cherry}
\affil[1]{School of Computational Science and Engineering, Georgia Institute of Technology, 756 West Peachtree Street, Atlanta, GA, 30308, United States}
\affil[2]{School of Physics, Georgia Institute of Technology, 837 State Street NW, Atlanta, GA, 30332, United States}
\affil[*]{These authors contributed equally to this work.}

\date{}

\maketitle

\begin{abstract}
Cardiac action potential models allow examination of a variety of cardiac dynamics, including how behavior may change under specific interventions. To study a specific scenario, including patient-specific cases, model parameter sets must be found that accurately reproduce the dynamics of interest. To facilitate this complex and time-consuming process, we present an interactive browser-based tool that uses the particle swarm optimization (PSO) algorithm implemented in JavaScript and taking advantage of the WebGL API for hardware acceleration. Our tool allows rapid customization and can find low-error fittings to user-provided voltage time series or action potential duration data from multiple cycle lengths in a few iterations (10-32), corresponding to a runtime of a few seconds on most machines. Additionally, our tool focuses on ease of use and flexibility, providing a webpage interface that allows users to select a subset of parameters to fit, set the range of values each parameter is allowed to assume, and control the PSO algorithm hyperparameters. We demonstrate our tool's utility by fitting a variety of models to different datasets, showing how convergence is affected by model choice, dataset properties, and PSO algorithmic settings, and explaining new insights gained about the physiological and dynamical roles of the model parameters.
\end{abstract}

\section{Background}

Mathematical models are widely used to predict and analyze the electrical behavior of cardiac cells and tissue. 
Although solving the differential equations that underlie most modern models of cardiac action potentials~\cite{fenton2008models} is a challenging numerical task~\autocite{clayton_models_2011}, due to the often large number of variables required to describe them as well as the large number of cells required for 2 and 3D simulations. 
Software like Chaste~\autocite{mirams_chaste_2013}, OpenCarp~\autocite{plank_opencarp_2021}, and a set of WebGL programs~\cite{kaboudian2019real,kaboudian2021real} have been developed to simplify obtaining model solutions.
However, even relatively simple cardiac models
include many parameters whose values can be modified to represent cells and tissue under different conditions and
to accommodate the inherent variability between individuals\cite{britton2013experimentally} 
Because these parameters have complex, interacting effects on the model output, 
finding a set of
parameters for a model that accurately captures the relevant behavior of data recorded from experiments or from
a different model is challenging.

A variety of approaches have been used to tackle the problem of finding appropriate parameter values, including simulated annealing~\autocite{lombardo2016comparison}, least squares~\autocite{dokos2004} and gradient-based approaches~\autocite{mathavan2009parameter}, 
and regression methods~\autocite{sarkar2010regression,tondel2014insight}. 
Genetic algorithms,
which maintain a pool of candidate solutions (parameterizations) that are improved 
using biologically-inspired
crossover, selection, and mutation transformations,
also have been widely used~\autocite{bot_rapid_2012,cairns2017,groenendaal2015,kherlopian2011cardiac,syed_atrial_2005}. 
Particle swarm optimization~(PSO) similarly maintains a pool
of candidate solutions, but rather than evolving the individuals according to rules inspired by genetics, the
candidates move in random degrees toward the best solutions that have been identified in previous
iterations~\autocite{chen2012identification,seemann2009adaption,rheaume2023modified}; 
%An advantage of the genetic and PSO algorithms is that they treat the model-solving step as a black box and will not terminate when a local optimum is found.  However, many candidates and iterations may be necessary to find an acceptable solution, especially when the dimension of the problem (number of parameters to fit) becomes large. 
it has also been combined with a quadratic optimization algorithm in a hybrid approach to improve convergence~\autocite{loewe2016parameter}. Techniques like data assimilation\cite{hoffman2016reconstructing,cherry_data_2017} also have been used to estimate parameter values along with the system state.
As an alternative to more traditional optimization methods, Bayesian approaches have been used to estimate not a single set of parameter values but
a posterior probability distribution---given a model output, how likely a particular model parameterization is---using
a variety of algorithms, such as Markov chain Monte Carlo~\autocite{siekmann2011}, approximate Bayesian
computation~\autocite{daly2015,daly2017}, Bayesian active learning~\autocite{zaman2021}, Bayesian history
matching~\autocite{coveney2018}, and Hamiltonian Monte Carlo~\autocite{nieto_quantifying_2021,nieto2023bayesian}. 
Implementations for many of these approaches to fitting cardiac model parameters may require extensive programming, so that the selection of a technique for a particular application may depend on prior experience and preferences.

In the present study, we present \ourtool, a tool to find cardiac model parameterizations that is fast, flexible,
easy to use, and applicable in many settings without requiring detailed understanding of the model parameters or optimization
algorithm. To promote computational efficiency, we 
utilize the parallelism provided by graphics hardware on modern computers, which influences our choice of optimization approach toward one that can benefit from this architecture.
Toward this end, \ourtool uses PSO
to optimize parameter values for several cardiac models to
match user-provided datasets.
PSO requires updating a pool
of candidate solutions over multiple 
iterations, with each
candidate solution at each iteration requiring a full solution of the model to be compared with the
data. Due to its limited communication needs, the PSO algorithm is well suited to be executed on thousands of cores in parallel.
Consequently, \ourtool can
implement PSO with a large candidate pool (up to tens of thousands) and often can identify a low-error parameter fitting within a few dozen iterations in seconds, even on modest hardware, without
sacrificing accuracy. As \ourtool uses the WebGL API for JavaScript and is implemented as a
webpage, no machine-specific manual compilation or installation is necessary to optimize
performance. The only setup required is visiting a website hosting the program or downloading the code and opening it in
a web broswer. The source code for the CardioFit program is publicly available on GitHub\footnote{\url{http://github.com/mcomstock/cardiofit-royalsociety-2024}}.

Previously, we have shown that this approach is capable of fitting a cardiac action potential (AP) model
to data taken from explanted human hearts with Brugada syndrome~\autocite{cairns2023automated} and to data generated from complex cardiac models~\autocite{rheaume2023modified}. In this
work, we present a detailed description of \ourtool's implementation and quantify its effectiveness. We find that \ourtool can fit datasets taken from biophysically complex models to several phenomenological and minimal cardiac models,
accurately capturing key properties of the original
data, such as AP duration (APD), alternans, and action potential morphology. We demonstrate similarly accurate fittings for a variety of
experimental data sets, with as many model parameter values being fit simultaneously as desired. We also show that with available graphics hardware, a large number of candidate solutions can be used to find
appropriate model parameterizations with just tens of PSO iterations. In addition, the fitting process leads to additional insights into the roles of parameters in cardiac models as well as identifiability. 
%, indicating that the parallel approach of \ourtool leads to its speed and efficiency in finding a low-error parameter set.

\section{Methods}

In this section, we provide information about the cardiac models and their simulation, the optimization algorithm, and the datasets considered as well as \ourtool's design and implementation. Additional details are available in the Supplement. 

\subsection{Cardiac action potential models}
\label{sec:models}
The two- to four-variable cardiac models used in \ourtool describe the change in a dimensionless voltage variable $u$ over time:
\begin{align}
    \frac{du}{dt} = I_\text{tot} + I_\text{stim},
\end{align}
where $I_\text{tot}$ is a function of $u$ and one or more variables that
evolve in time according to the specific model and $I_\text{stim}$ is an external stimulus current. The models currently included in \ourtool are the modified FitzHugh-Nagumo (MFHN)~\autocite{velasco2022methods}, Mitchell-Schaeffer (MS)~\autocite{mitchell2003}, modified Mitchell-Schaeffer (MMS)~\autocite{corrado2016two}, Fenton-Karma~\autocite{fenton1998}, Bueno-Orovio-Cherry-Fenton (BOCF)~\autocite{bueno2008}, and Brugada BOCF (BBOCF) models~\autocite{bueno2015basis}. Detailed equations for each model are supplied in the Supplement.
Models are integrated using
the forward Euler method with a time step of \qty{0.02}{\milli\second}.

Two types of stimulus current are available to the user. The default option is a square stimulus, which is set to a non-zero constant for a fixed amount of time and is zero otherwise.
The default magnitude of the stimulus is \qty{0.2}{\per\milli\second} and the default duration is \qty{2}{\milli\second}. These values can be changed in the user interface, as described in Section~\ref{sup:sec:interface-stimulus}.
Additionally, a biphasic stimulus designed to mimic the upstroke-inducing current experienced via diffusive coupling is available. Details are given in the Supplement. 

\subsection{Experimental and model-derived datasets}

% Please provide information on the datasets used: species, microelectrode or optical if known (I think these are mostly microelectrode except for human), which cycle lengths are available/you are using, time resolution. You can indicate that the data were obtained under pacing protocols (e.g., pace to a steady state for each CL). Flavio and I can add any necessary information on how the datasets were obtained.

% Cycle lengths used in ms
% MS: 500 400 300
% FK: 500 400 300
% BOCF: 500 400 300
% Fox: 500 400 300 225
% TP: 500 400 350 310
% Canine: 500 400 320 260
% Fish: 300
% Frog: 800
% Human Brugada: 1000

% Time resolutions in ms - 1 in all cases

% Not sure if fish and frog data are microelectrode or optical mapping - we need to ask Darby or Flavio.
% Human Brugada is optical
% All others are microelectrode

% Anything else about the data, we need to ask someone else

To demonstrate the versatility of \ourtool, we show fittings to both experimental and model-derived data.  Our experimental datasets include recordings from four different
datasets: fish, frog, canine endocardial, and human. The fish, frog, and canine datasets are from
mircroelectrode recordings from a single ventricular cell in tissue paced at one or more constant periods, or cycle lengths (CLs). 
The human ventricular dataset consists of time series taken from single
pixels of optical-mapping recordings of an explanted human heart displaying characteristics of Brugada
syndrome. The time resolution of all experimental datasets is \qty{1}{\milli\second}.
Datasets used were recorded with cycle lengths \qty{500}{\milli\second}, \qty{400}{\milli\second}, \qty{320}{\milli\second}, and \qty{260}{\milli\second} for canine data,
\qty{300}{\milli\second} for fish data, \qty{800}{\milli\second} for frog data, and \qty{1000}{\milli\second} for human data.

Model-derived data were generated from any of the models available within \ourtool and from more complex cardiac action potential models, specifically,
the Fox et al.\ model~\autocite{fox2002ionic} and the ten Tusscher et al.\ model~\autocite{ten2006alternans}. These models were integrated using the Rush-Larsen method for gating variables and forward Euler for other variables with a time step of \qty{0.005}{\milli\second}, and
the corresponding datasets all have a time resolution of \qty{1}{\milli\second}. Calcium concentrations in the ten Tusscher et al.\ model are updated using the analytical approach described in Appendix~1 of Ref.~\autocite{zeng1995two}. Pacing cycle lengths of \qty{500}{\milli\second}, \qty{400}{\milli\second}, and \qty{300}{\milli\second} were used
for each of the MS, FK, and BOCF models. For the Fox. et al. model dataset, pacing cycle lengths of \qty{500}{\milli\second}, \qty{400}{\milli\second}, \qty{300}{\milli\second}, and \qty{225}{\milli\second} were used. Data with cycle lengths of \qty{500}{\milli\second}, \qty{400}{\milli\second}, \qty{350}{\milli\second}, and \qty{310}{\milli\second} are used from the
ten Tusscher et al.\ model. In all cases, the model was paced until a steady state was reached for the given cycle length before recording the data.

\subsection{Particle swarm optimization algorithm and \ourtool implementation}
\label{sec:pso-algorithm}

To identify appropriate parameter values that fit data to any of the models listed in Sec.~\ref{sec:models}, \ourtool uses PSO, a derivative-free optimization method that does not require specific assumptions about
the problem being optimized. It maintains a pool of candidate solutions
(``particles'') that are iteratively updated, using knowledge from the previous particle
states, until a fixed number of iterations is completed (PSO can also be implemented to terminate when some threshold criterion is achieved,
or the particles converge to a stable result).
Each particle has a position and velocity with $d$~dimensions, where $d$ is the number of parameters
of the model. The position $\vec{p}_i$ of particle $i$ is a vector consisting of the current values of each of the parameters
for that particle, and the velocity is used to determine changes in the position according to
information acquired from the group of particles. Information
used to update the velocity of a particle $i$ includes the best position it has ever achieved, $\vec{b}_i$,
and the best position ever found by any particle, $\vec{b}$.

%\subsubsection{\ourtool PSO Settings}
For fitting cardiac models to voltage data, \ourtool uses the sum of
squared error between the normalized data and the voltage of the model as the fitness metric.
The program automatically aligns the first upstroke of the model output and the data to perform the comparison.
\ourtool also can be used to fit the model to action potential duration (APD) values in addition to or instead of a
full voltage time series, in which case the absolute APD error is used. When both types of error are considered
or multiple data sets are provided, an additional parameter can be adjusted to balance their relative weights. The
error is automatically weighted by the length of the data file so that longer data sets (corresponding to longer CLs) do not necessarily tend
toward higher error. For a given dataset, the quality of a particular particle is determined by the value
of the fitness function evaluated using its parameter values.

The position of each particle is initialized from a uniform random distribution over the valid range of values
for each parameter. The position and velocity of each particle are
updated at each iteration $n$ according to
\begin{align}
    \vec{v}_i^{n+1} &= 
    \begin{aligned}[t]
        \chi [\vec{v}_i^n + &\vec{u}(0,\phi_1) \otimes \left(\vec{b}_i^n-\vec{p}_i^n\right)\\
        + &\vec{u}(0,\phi_2) \otimes \left(\vec{b}^n-\vec{p}_i^n\right)],
    \end{aligned}\\
    \vec{p}_i^{n+1} &= \vec{p}_i^n + \gamma \vec{v}_i^{n+1},
\end{align}
where $\chi$ is a constriction coeffecient used to promote convergence in the
algorithm~\autocite{clerc2002particle}, $\gamma$ is the learning rate that governs how aggressively particle positions are changed, $\vec{u}(a,b)$ is a $d$-dimensional vector of uniformly distributed random
numbers in the range $[a,b)$, and $\otimes$ is the Hadamard product. Each
parameter is restricted to a range of possible values; if an updated particle contains any parameter value
outside its range, that value is reset to a random position within the nearest three-quarters of
the range. The values of $\phi_1$, $\phi_2$, and $\chi$ are hyperparameters that may be 
adjusted as desired using the user interface as described in Section~\ref{sup:sec:interface-stimulus}. The results presented here use \ourtool's default values of $\phi_1=\phi_2=2.05$ and
$\chi = 2 / (\phi-2+\sqrt{\phi^2-4\phi}) \approx 0.73$
with $\phi=\phi_1+\phi_2$~\autocite{loewe2016parameter}, as well as $\gamma=0.05$, which was determined empirically.

The PSO algorithm relies on a pool of particles to explore parameter space and
find viable solutions. Because the evaluation of each particle requires finding the solution to the model
and comparing the result with each value of the input data to fit, each iteration of the algorithm
incurs a computational effort proportional to the number of particles. However, these model runs require no interaction between
particles, and therefore may be performed in parallel with no modification to the algorithm. Other
steps, such as the particle update step, may also be performed in parallel to further
increase the speed of the program and avoid unnecessary data transfer. More details are available in Section~\ref{sup:sec:parallel-impl}.

\subsection{\ourtool Interface}
\label{sec:interface}

The user interface of \ourtool provides access to the key features 
of our PSO implementation without requiring knowledge of the underlying implementation. The following sections
contain a brief description of the provided behavior and intended workflow for the program. Screenshots of the
main user interface and additional interface elements of \ourtool are shown in Figures~\ref{fig:full-ui} and~\ref{fig:ui-more} for reference. More details are provided in the Supplement.

\begin{figure*}[t]
    \centering
    \includegraphics[width=129mm]{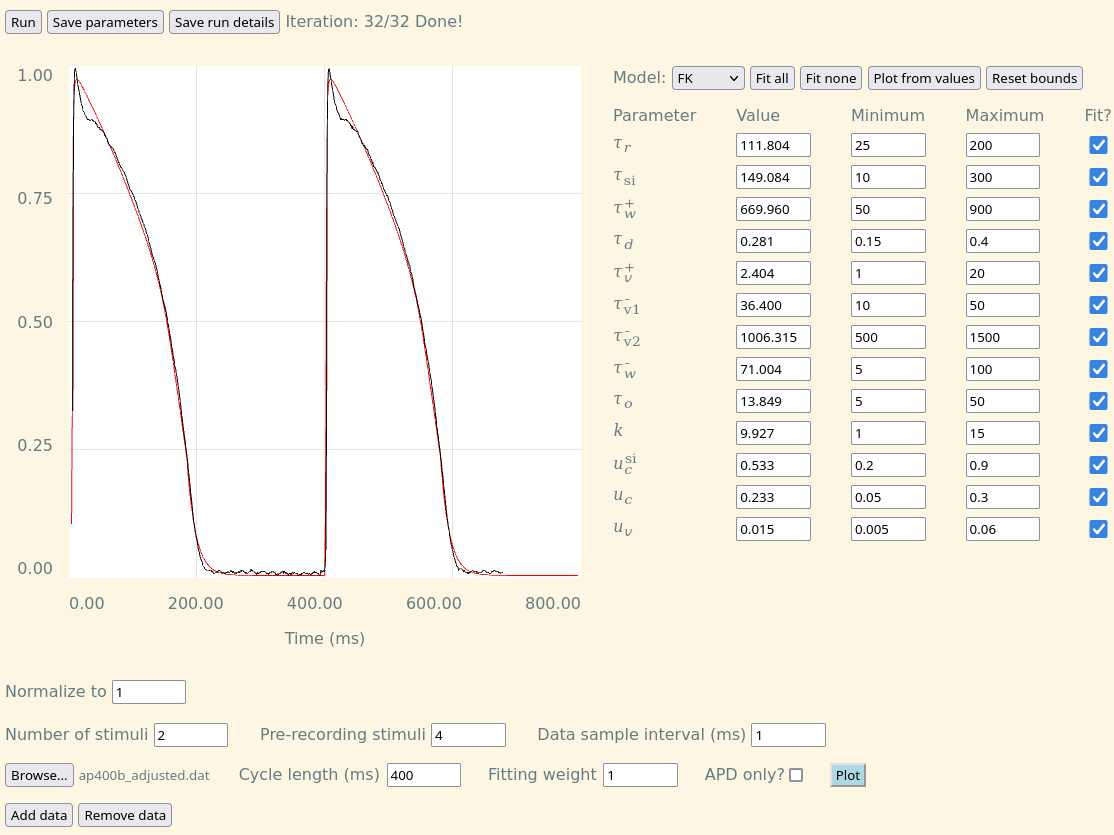}
    \caption{User interface of \ourtool. Top row: Buttons to initiate and save the results of
    \ourtool runs and a progress display for the iterations. Left: Fit (red) to the data (black) for the selected cycle length. Right: Interface for choosing the model, selecting the parameters to fit, setting the bounds, and viewing the
    fit parameter values. Bottom: Interface for adding datasets to fit.}
    \label{fig:full-ui}
\end{figure*}

\subsubsection{Specifying data to fit}

Fitting a model to a dataset in \ourtool requires adding the data and setting a few options; see Figure~\ref{fig:apd-ui}. \ourtool expects
newline-delimited files containing a sequence of voltage values. The data are normalized by the program to have a minimum
value of zero and a maximum value matching the number entered in the ``Normalize to'' entry field. If the dataset
is already normalized to the desired range, this behavior can be bypassed by setting the value of the
``Normalize to'' field to $0$. Default values for the ``Normalize to'' field vary by model and are given in Section~\ref{sup:sec:model-eqns}; these values are used in the results presented here unless noted otherwise. Note that within the figures all results have been normalized to the same 0-1 scale, even if a normalization constant other than one was used for a fitting.

Aside from normalization, other pre-processing steps outside of \ourtool may be necessary. For
instance, high-resolution data should be downsampled to a sample interval of around
\qtyrange{1}{2}{\milli\second}, as large data files may not fit in GPU memory on systems without powerful dedicated graphics hardware. When
comparing the model output with a candidate parameterization to the provided data, the model is first run for
a user-specified number of stimuli, set in the ``Pre-recording stimuli'' field, to remove transient effects of the initial conditions. After these stimuli have been applied, the simulation continues by applying the number of stimuli specified as the ``Number of stimuli''
field, and the results are compared with the data.

To add a single data file, the ``Browse'' button above the ``Add data'' button is clicked and the file is selected. Along
with this file, the pacing period is entered in the ``Cycle length'' field to the right. Additional files with
different pacing periods can be added by clicking the ``Add data'' button to add additional rows for data
entry. These rows may be removed by clicking the ``Remove data'' button. The ``Fitting weight'' field allows the
different files to be given greater or lesser relative priority in the PSO fitness function when multiple files are used.
When the total error for a particular particle is computed, the error from each dataset is multiplied by its fitting
weight, and then all the components are summed. It is recommended to
leave the fitting weights at the default value of 1 for initial attempts and adjust them to fine-tune the behavior of
subsequent fits. 
For example, data from a particular cycle length could be given extra weight in an attempt to fit a bifurcation or alternans details. 
%For example, if the \ourtool result seems to be ignoring a particular dataset, this problem can potentially be corrected
%by increasing the corresponding fitting weight. Similarly, if a dataset is having too much influence on the fit, decreasing its fitting weight may
%solve the problem. 
Additionally,
the number of stimuli to apply, number of pre-recording stimuli to be applied before fitting, and sample
interval of the data must be the same for all data files and entered above the data rows. Larger numbers of
stimuli and smaller sample intervals will increase the time required for the program to run.

The ``Plot'' button in each data row can be used after selecting a file to plot the corresponding raw data as a time series.
After running \ourtool, the functionality of this button changes to plot both the data and the model fit.

If only APD data is available, then the ``APD only'' checkbox gives the option to fit to APDs. When the
checkbox is selected, the data row changes to contain the entries shown in Figure~\ref{fig:apd-ui}. Instead
of selecting a file, the APD values are entered directly in the ``APDs'' field as a comma-separated list of
numbers. The number of APDs entered in the ``APDs'' field should match the value in the ``Number of stimuli''
field above, so that one correct value is provided for comparison with each simulated action potential. The
``APD threshold'' corresponds to the normalized voltage value at which the APD is measured: e.g., $\text{APD}_{90}$ for
data normalized to \num{1} would be specified by entering \num{0.1}.

It is possible to use \ourtool to fit voltage and APD data simultaneously, in which case the ``Fitting weight'' field becomes relevant to tune the relative importance of the datasets.
To fit both voltage and APD data, multiple data rows must be added as described above, with separate rows representing the voltage
data and the APD data.

\def\imagetop#1{\vtop{\null\hbox{#1}}}

\begin{figure}
    \centering
    \begin{subcaptiongroup}
    \begin{tabular}{l l}
        \imagetop{A} & \imagetop{\includegraphics[width=129mm]{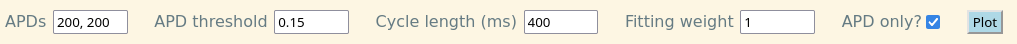}}
        \phantomcaption{}
        \label{fig:apd-ui}\\
        \imagetop{B} & \imagetop{\includegraphics[width=84mm]{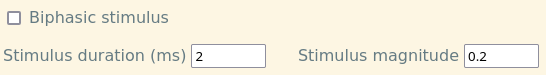}}
        \phantomcaption{}
        \label{fig:squarestim-ui}\\
        \imagetop{C} & \imagetop{\includegraphics[width=129mm]{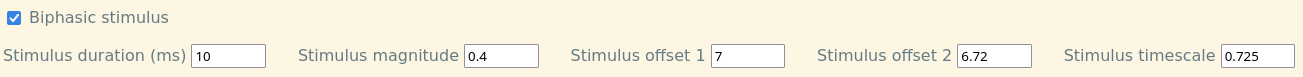}}
        \phantomcaption{}
        \label{fig:biphasicstim-ui}\\
        \imagetop{D} & \imagetop{\includegraphics[width=84mm]{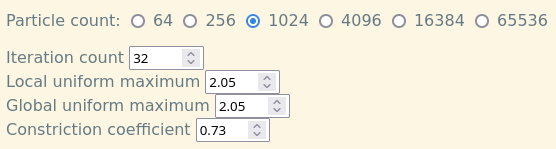}}
        \phantomcaption{}
        \label{fig:ui-hyperparams}
    \end{tabular}
    \end{subcaptiongroup}
    \caption{Detailed view of several elements of the \ourtool user interface. (a) APD data entry interface for \ourtool. 
    %The ``APDs'' field is a comma-separated list of APDs to fit; the ``APD threshold'' field is the normalized voltage value to use for measuring APDs; the``Cycle length'' field is the pacing cycle length, and the ``Fitting weight'' field is used to set therelative importance of fitting this data line within the PSO algorithm.
        (b) Square stimulus parameters for \ourtool.
        (c) Biphasic stimulus parameters for \ourtool.
        (d) The interface to configure the PSO algorithm hyperparameters in \ourtool.}
        \label{fig:ui-more}
\end{figure}

%\begin{figure}
%    \captionsetup[subfigure]{position=top,labelfont=bf,textfont=normalfont,singlelinecheck=off,justification=raggedright}
%    \centering
%    \begin{subfigure}{\textwidth}
%%        \centering
%        \caption{}
%        \includegraphics[width=129mm]{Figures/ui_apds.png}
%        \label{fig:apd-ui}
%    \end{subfigure}
%    \begin{subfigure}{\textwidth}
%%        \centering
%        \caption{}
%        \includegraphics[width=84mm]{Figures/ui_squarestim.png}
%        \label{fig:squarestim-ui}
%    \end{subfigure}
%    \begin{subfigure}{\textwidth}
%%        \centering
%        \caption{}
%        \includegraphics[width=129mm]{Figures/ui_biphasicstim.png}
%        \label{fig:biphasicstim-ui}
%    \end{subfigure}
%    \begin{subfigure}{\textwidth}
%%        \centering
%        \caption{}
%        \includegraphics[width=84mm]{Figures/ui_hyperparams.png}
%        \label{fig:ui-hyperparams}
%    \end{subfigure}
%    \caption{Detailed view of several elements of the \ourtool user interface. (a) APD data entry interface for \ourtool. 
%    %The ``APDs'' field is a comma-separated list of APDs to fit; the ``APD threshold'' field is the normalized voltage value to use for measuring APDs; the``Cycle length'' field is the pacing cycle length, and the ``Fitting weight'' field is used to set therelative importance of fitting this data line within the PSO algorithm.
%        (b) Square stimulus parameters for \ourtool.
%        (c) Biphasic stimulus parameters for \ourtool.
%        (d) The interface to configure the PSO algorithm hyperparameters in \ourtool.}
%\end{figure}

\subsubsection{Choosing a model and parameter ranges}

The fields to the right of the graph window are used to select the model whose parameters will be fit and to
adjust model parameter settings. Once a model is selected by clicking the ``Model'' drop-down menu
and choosing one of the options, the other fields automatically change to
display the relevant parameters and default values. Each row beneath the model name contains, from left to
right, the parameter name, the value of that parameter (initially empty), the minimum allowable value of the
parameter for the fitting, the maximum allowable value of the parameter for the fitting, and a checkbox
indicating whether or not that parameter is to be fit. The ``Fit all'' and ``Fit none'' buttons provide
convenient shortcuts to (respectively) check or uncheck all of these boxes. If a parameter is not set to be
fit, a value must be provided in the ``Value'' column. For all other parameters, the range of values
explored by PSO can be adjusted by changing the maximum and minimum values. All results presented here use the default parameter bounds for each model (see Tables~\ref{sup:tab:mfhn-bounds}--\ref{sup:tab:bbocf-bounds}) unless stated otherwise.

To generate a fit using \ourtool, click the ``Run'' button at the top of the page. The
current iteration number is displayed while the program runs. Once the program has completed, the
result of the fit compared with the first data file is  plotted in the graph window, as shown in
Figure~\ref{fig:full-ui}, with the model fit in red and the data in black. A comparison of the output fit for
a different input data file may be viewed by clicking the ``Plot'' button next to the file of interest. A plot of the global best
error (vertical axis) as the number of iterations increases (horizontal axis), as shown in Figure~\ref{sup:fig:ui-error-plot},  is plotted beneath the PSO hyperparameters.

\subsubsection{Interpreting and saving the results}

After the iterations are completed, the best-fit value resulting from the fitting is displayed in the ``Value'' column for
each parameter. Note that the displayed values indicate the actual values and precision of the parameters; three decimal places of precision are used. The parameter values can be saved to a file by clicking the
``Save parameters'' button. The parameter values and other additional
information about the \ourtool run, such as the PSO hyperparameter values, can be saved to a file by clicking the
``Save run details'' button. Because \ourtool runs in a web browser, the file is treated as a download;
it may appear in the ``Downloads'' folder, or a prompt to select a save location may appear,
depending on the browser settings. 

\subsection{Statistical analysis}

For comparisons of parameter values obtained across multiple fittings of a given model to different datasets, we used the standard t-test. Because of the large number of model parameters and dataset comparisons, we used a small p-value of 0.001 to determine significance, but we also identified cases with p-values between 0.001 and 0.01. For all comparison cases, we used 20 runs with \num{4096} particles and \num{32} iterations.

\section{Results}

In this section, we demonstrate the effectiveness of \ourtool in  fitting both model-generated data and experimental data and highlight how \ourtool can lead to insights about the roles of the parameters in the models being fit. We also analyze \ourtool's convergence and scaling properties.

\subsection{Fitting to data from the same model}
\label{sec:model-recovery}

\begin{figure}[htpb]
    \centering
    \includegraphics{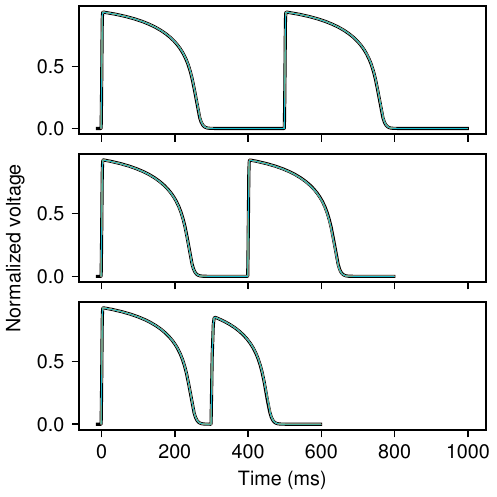}
    \caption{MS model fit to voltage data obtained from the same model. Cycle lengths of
    \qty{500}{ms}, \qty{400}{ms}, and \qty{300}{ms} were fit simultaneously to the two action potentials shown. Reference data is plotted in black; the
    results of \num{20} separate fits using \ourtool are plotted in various colors. Fits were generated using \num{4096}
    particles, \num{100} iterations, and four pre-recording stimuli in all cases.}
    \label{fig:self-3cl-ms}
\end{figure}

\begin{figure}[htpb]
    \centering
    \includegraphics{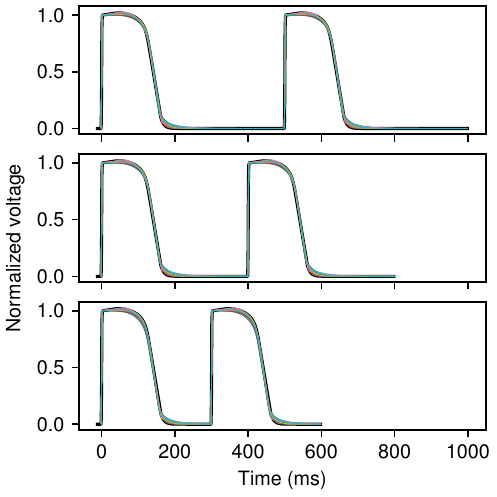}
    \caption{FK model fit to voltage data obtained from the same model. Cycle lengths of
    500, 400, and \qty{300}{ms} were fit simultaneously to the two action potentials shown. Reference data is plotted in black; the
    results of \num{20} separate fits using \ourtool are plotted in various colors. Fits were generated using \num{4096}
    particles, \num{100} iterations, and four pre-recording stimuli in all cases.}
    \label{fig:self-3cl-fk}
\end{figure}

\begin{figure}[htpb]
    \centering
    \includegraphics{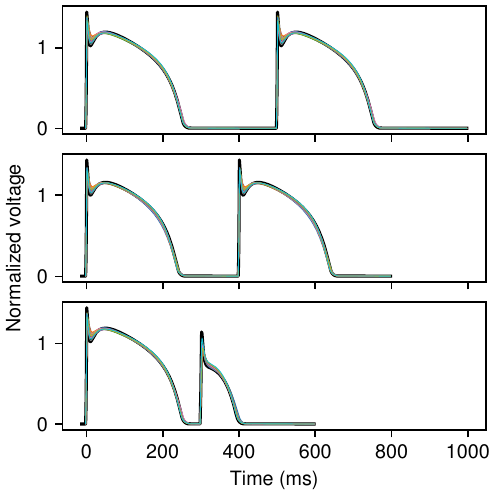}
    \caption{BOCF model fit to voltage data obtained from the same model. Cycle lengths of
    500, 400, and \qty{300}{ms} were fit simultaneously to the two action potentials shown. Reference data is plotted in black; the
    results of \num{20} separate fits using \ourtool are plotted in various colors. Fits were generated using \num{4096}
    particles, \num{100} iterations, and four pre-recording stimuli in all cases.}
    \label{fig:self-3cl-bocf}
\end{figure}

A first test of \ourtool is fitting a model to data derived from the same model, which allows us not only to evaluate the quality of the fit (how close the voltage and/or APD values are to the user-provided data), but also to assess how reliably the known parameters used to create the model-generated dataset can be identified from the data. 
Due to randomness in the particle initialization and velocity updates, combined with limitations in parameter identifiability from the model structure or dataset(s) being fit (or both), repeated fitting attempts of a model to data generated from the same model may result in different parameter values that result in slightly different voltage traces. 
We tested this behavior with the MS, FK, and
BOCF models, which were fit to three cycle lengths simultaneously. 
Figures~\ref{fig:self-3cl-ms}--\ref{fig:self-3cl-bocf} show 
the data from
the true model parameterization in black and the result of \num{20} different fittings overlaid in
color for the three models. Because the MS model has only five
parameters, the \num{20} fittings produced results that are difficult to distinguish visibly, as shown in Figure~\ref{fig:self-3cl-ms}. For the FK model, variability in the fitted voltage values obtained occurred during the AP plateau and toward the end of repolarization, as can be seen in
Figure~\ref{fig:self-3cl-fk}.  Figure~\ref{fig:self-3cl-bocf} shows that for the BOCF model, differences were evident primarily in the height of the upstroke as well as during the initial repolarization and plateau.

\begin{figure*}
    \centering
    \includegraphics{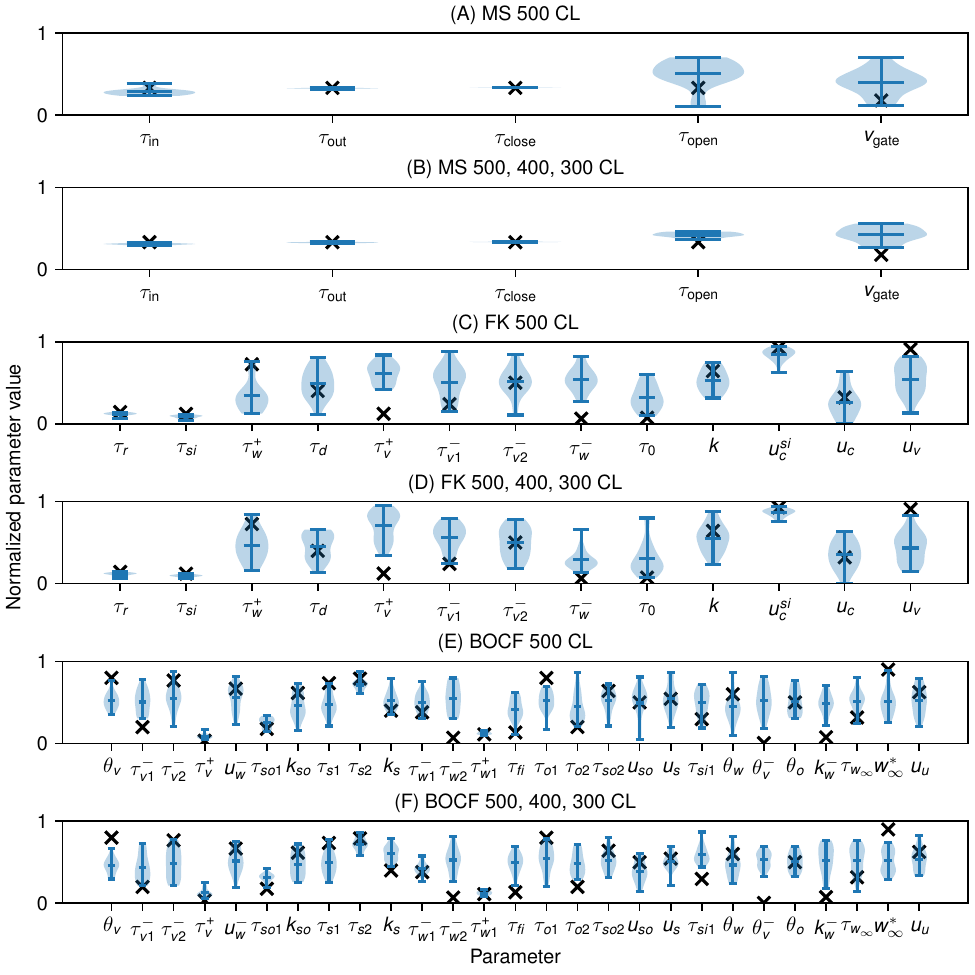}
    \caption{Violin plot of parameter values from the BOCF, FK, and MS models fit to data
    taken from the respective model itself over 20 runs. The black ``X'' indicates the value of the
    parameter used to generate the data. The parameter values are presented normalized over
    the range of parameter bounds, using the defaults in all cases.
    (A) MS model fit to itself, with one dataset with a cycle length of \qty{500}{\milli\second}.
    (B) MS model fit to itself, with three datasets, cycle lengths 300,
    400, and \qty{500}{\milli\second}.
    (C) FK model fit to itself, with one dataset with a cycle length of \qty{500}{\milli\second}.
    (D) FK model fit to itself, with three datasets, cycle lengths 300,
    400, and \qty{500}{\milli\second}.
    (E) BOCF model fit to itself, with one dataset with a cycle length of \qty{500}{\milli\second}.
    (F) BOCF model fit to itself, with three datasets, cycle lengths 300,
    400, and \qty{500}{\milli\second}. Fits were generated using \num{4096}
    particles, \num{100} iterations, and four pre-recording stimuli in all cases.}
    \label{fig:self-recovery}
\end{figure*}

Because the datasets came from models with known parameter values, these examples also allowed us to examine the accuracy of the parameter values found as well as their variability. Figure~\ref{fig:self-recovery}~(A--B) shows that for the MS model, 
two of the parameters, $\tau_{close}$ and $\tau_{out}$ were identified consistently near the true values across the 20 fittings, while the others showed more variability. The choice of datasets used in the fitting could affect the ability to recover the true parameter values, as shown in Figure~\ref{fig:self-recovery}~(B), where including more cycle lengths resulted in identifiability improvements in the form of reduced standard deviations for $\tau_{in}$ and, to a lesser degree, for $\tau_{open}$ and $v_{gate}$. 

Fitting the increased number of parameters for the FK and BOCF models made finding a unique minimizing set of parameter values more challenging, as can be seen in Figure~\ref{fig:self-recovery}~(C--F).
In some cases, such as $\tau_v^+$ for the FK model and $\tau_{w2}^-$ for the BOCF model, the range of values obtained for a parameter across the 20 fittings did not even include the original value. 
Although fitting to more CLs decreased the variability in some of the parameter values obtained using \ourtool (8 of the 13 for the FK model and 18 of the 27 for the BOCF model had lower standard deviations when fit with multiple CLs), it did not guarantee better constraints on all parameters, with many parameter ranges remaining the same and some even increasing (such as $k$ for the FK model and $k_w^-$ for the BOCF model). 
This sustained variability in many parameters suggests complex interactions among the parameters along with a limited ability to recover parameters governing the gating variables with only voltage datasets for these models.
Nevertheless, a small number of parameters could be identified consistently, including $\tau_r$, $\tau_{si}$, and $u_c^{si}$ 
for the FK model and $\tau_v^+$, $\tau_{so1}$, and $\tau_{w1}^+$ for the BOCF model. Such parameters represent values to which the model is highly sensitive. For example, $\tau_r$ in the FK model directly governs the strength of the repolarizing current, so that errors in this value noticeably alter the action potential shape.

%We call attention to the fact that similar waveforms do not necessarily indicate similar parameterizations. In
%fact, even though all fits produced waveforms that are quite similar to the data, much more variance was found in the
%corresponding parameters. This behavior is discussed further in Section~\ref{sec:parameter-variance}.

\subsection{Fitting to data from other models}
\label{sec:model-fit}

\begin{figure}[htpb]
    \centering
    \includegraphics{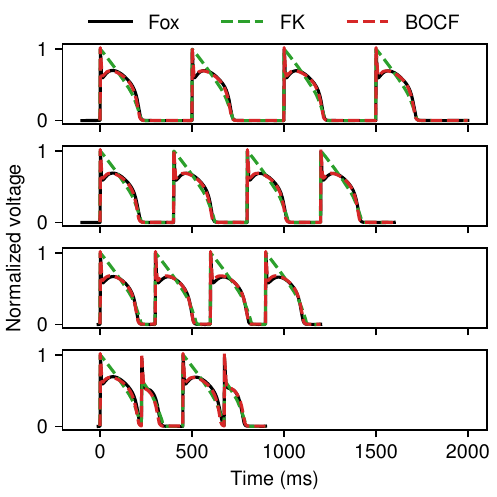}
    \caption{Fitting the FK (blue) and BOCF (orange) models to normalized voltage data obtained from the Fox et. al model (black). Action potentials at CLs of 500, 400, 300, and \qty{225}{ms} were fit simultaneously to the data shown, with alternans present at the shortest CL. Four pre-recording stimuli, \num{65536} particles, and \num{100} iterations were used in all cases, and the normalization constant was set to 1.4 for the BOCF model.}
    \label{fig:fox}
\end{figure}

\begin{figure}[htpb]
    \centering
    \includegraphics{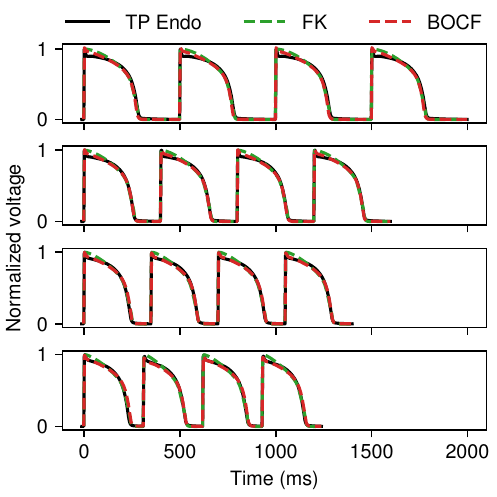}
    \caption{Fitting the FK (blue) and BOCF (orange) models to normalized voltage data obtained from the ten Tusscher-Panfilov model with endocardial tissue parameters (black). Action potentials at CLs of 500, 400, 350, and \qty{310}{ms} were fit simultaneously to the data shown. Four pre-recording stimuli, \num{65536} particles, and \num{100} iterations were used in all cases, and the normalization constant was set to 1.3 for the BOCF model.}
    \label{fig:tp_endo}
\end{figure}

To demonstrate the versatility of \ourtool, we used it to fit available models to data generated from
biophysically detailed models. The resulting parameterization may allow a less computationally expensive model to
reproduce aspects of the behavior of a more complex model.
Figure~\ref{fig:fox} shows results from fitting the FK and BOCF models to data from the 13-variable Fox et al. canine ventricular action potential model~\autocite{fox2002ionic}. Normalized voltage traces from four different cycle
lengths (500, 400, 300, and \qty{225}{ms}) were fit simultaneously, with the \qty{225}{ms}
CL producing large-amplitude alternans. 
As can be seen
in Figure~\ref{fig:fox}, the
FK model (blue) could not match the spike-and-dome AP shape of the Fox et al. model, but it nevertheless matched the peak voltage value and repolarization phases, while achieving alternans at the shortest CL and avoiding it for the longer CLs.
Using the BOCF model (orange), the Fox et al. data were reproduced more closely over the full APs for all CLs, including during alternans.

As another example of using \ourtool to fit data from a more complex model, Figure~\ref{fig:tp_endo} shows fits of the FK (blue) and BOCF (orange) models to normalized endocardial action potentials from the human
ventricular model of ten Tusscher et al.~\autocite{ten2006alternans}. In this case, the four CLs used were
500, 400, 350, and \qty{310}{ms}, with alternans occurring for the shortest CL.
Good agreement was obtained for both the FK and BOCF models with the AP shapes as well as the presence
or absence of alternans. The BOCF model more closely fit the initial repolarization and AP plateau than the FK model due to its greater flexibility in AP shape.

\begin{figure*}[htpb]
    \centering
    \includegraphics{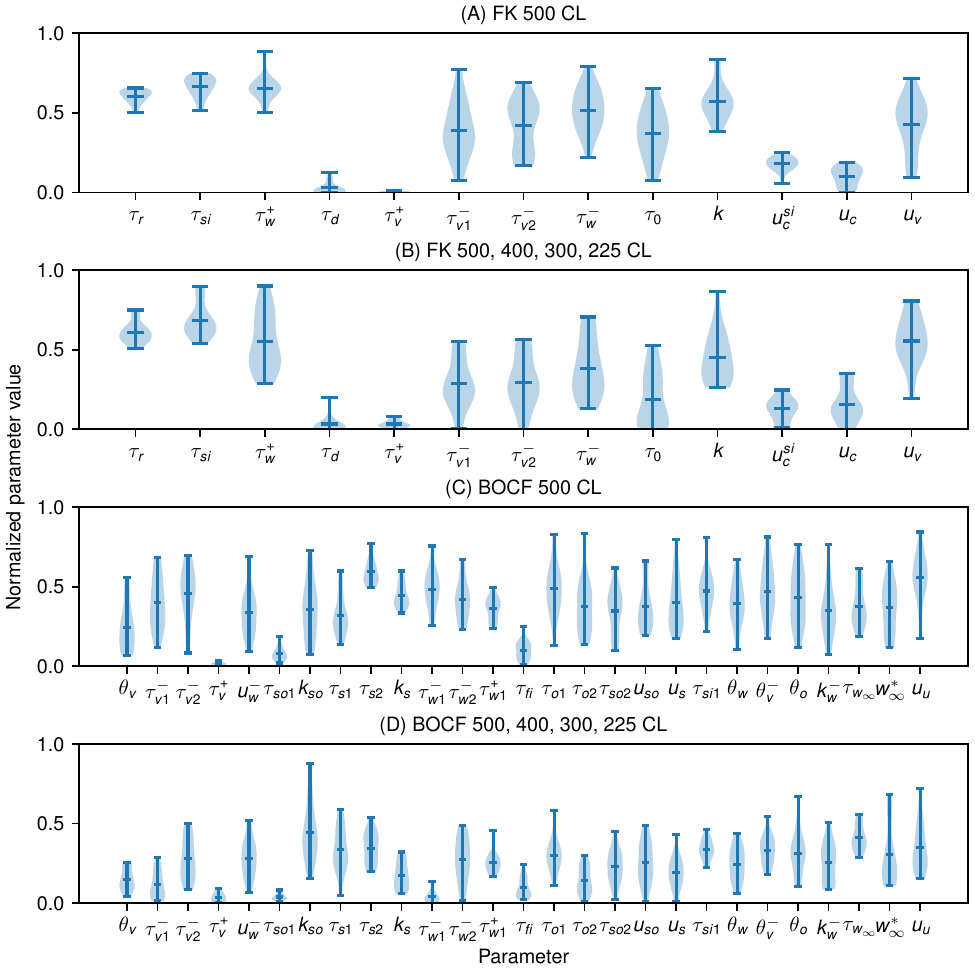}
    \caption{Violin plot of the parameter values of the BOCF and FK models taken over 20
    runs using the data from
    the Fox et. al model shown in Figure~\ref{fig:fox}. Parameter values are presented
    as normalized over the range of parameter bounds (the default bounds were used in all cases). 
    (A-B): Parameter values obtained for the FK model fit to the
    \qty{500}{\milli\second} data only (A) and to all four CLs (B). 
    (C-D): Parameter values obtained for the BOCF model fit to the \qty{500}{\milli\second} CL data only (C) and to all four data sets (D). All cases used
    \num{4096} particles and \num{100} iterations, and the normalization constant was set to 1.4 for the BOCF model.}
    \label{fig:parameter-variance}
\end{figure*}

Although there is no true set of parameters for the simplified models in \ourtool when fitting them to data obtained from other models, an examination of the consistency of the parameter values found can reveal the relative importance of the parameters of the simpler models as well as the usefulness of fitting data from multiple CLs. Figure~\ref{fig:parameter-variance} shows the normalized values of each parameter within
its default bounds over \num{20} runs using the data from the Fox et. al model described in
Section~\ref{sec:model-fit} to fit the FK and BOCF models.
Although the values for a few parameters for each model occupy only a small portion of the range defined by their bounds, many of the parameters can take on values over much of their ranges in parameterizations that give low-error fits to the voltage data. In general, the FK model parameters seemed more adequately constrained than the parameters of the BOCF model, which were more numerous by about a factor of two. In particular, the values of $\tau_d$ and $\tau_v^+$ were among the most consistent for the FK model.
In addition, increasing the number of cycle lengths in the data to be fit had different effects for the two models, with standard deviations decreasing for only two parameters for the FK model, as shown in panel~B as compared to~A; for the BOCF model, additional CLs reduced parameter ranges more appreciably, with standard deviations decreasing for 18 parameters, as shown in panel~D as compared to~C.
%Therefore, when certain parameter values or ranges are desired in \ourtool's output, they should
%be specified using the interface described in Section~\ref{sec:interface}.

\subsection{Fitting experimental data}

\begin{figure}[htpb]
    \centering
    \includegraphics{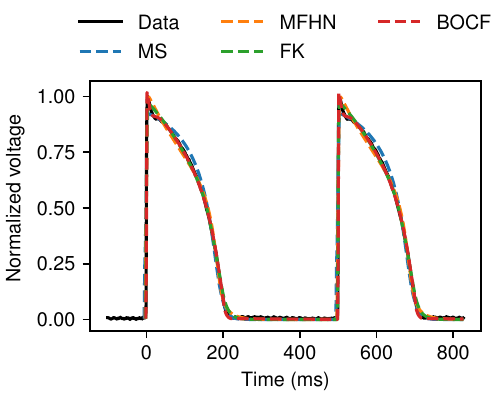}
    \caption{Fitting the MS (blue), MFHN (orange), FK (green), and BOCF (red) models to canine endocardial data paced at \qty{500}{\milli\second}
    using \num{4096} particles, \num{32} iterations, and four pre-recording stimuli, and the normalization constant was set to 1.35 for the BOCF model. Excellent agreement is obtained in all cases, with only small discrepancies shortly after the upstroke.}
    \label{fig:4fit}
\end{figure}

\begin{figure}[htpb]
    \centering
    \includegraphics{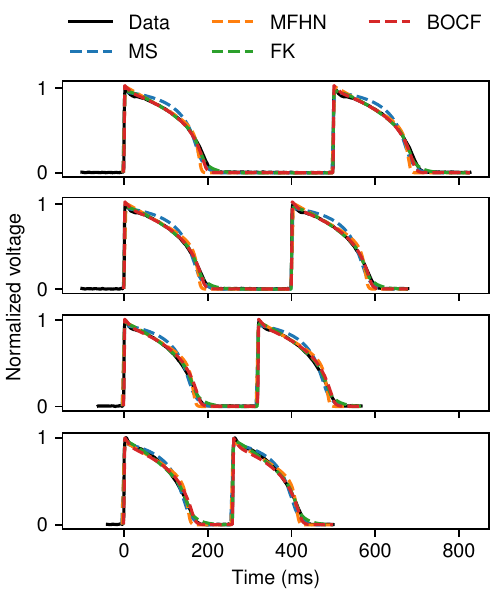}
    \caption{Fitting the MS, FK, and BOCF models to four sets of canine endocardial data paced at
    500, 400, 320, and
    \qty{260}{\milli\second}, using \num{4096} particles, \num{32} iterations, and four pre-recording stimuli, and the normalization constant was set to 1.35 for the BOCF model. The FK and BOCF models achieve slightly better fits than the MS model, which tends to produce action potentials that are less triangular than the APs in this dataset.}
    \label{fig:canine_fit_many_multicl}
\end{figure}

We also used \ourtool to fit models to microelectrode recordings of canine endocardial action potentials. Fits of the MS,
MFHN, FK, and BOCF models to normalized voltage data from a single CL of \qty{500}{\milli\second}
are shown in Figure~\ref{fig:4fit}. Very good agreement was obtained in all cases, with discrepancies most noticeable at and shortly after reaching the peak voltage value, due to differences in the models' characteristic action potential shapes. 
Figure~\ref{fig:canine_fit_many_multicl} shows the more challenging case of simultaneously fitting data from four different cycle
lengths of 500, 400, 320, and
\qty{260}{\milli\second} using the MS, FK, and BOCF models.
%Each model fitting used 32 iterations of the PSO algorithm with \num{1024}
%particles. 
In all cases, \ourtool identified a parameterization of
the model that closely matched the data within the limitations of the model, with a slightly more noticeable discrepancy for the MS model compared to the FK and BOCF models.

% FK: 5.347 seconds
% MS: 4.527 seconds
% Modified FHN: 4.5 seconds
% Bueno-orovio: 5.835 seconds

\begin{figure}[htpb]
    \centering
    \includegraphics{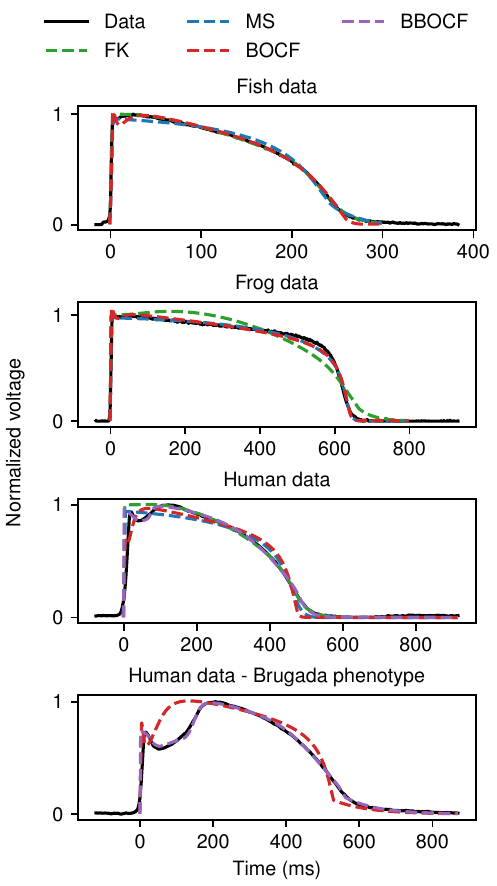}
    \caption{Representative fittings of the FK (blue), MS (orange), BOCF (green), and BBOCF (red) models to experimental data
        recorded from three different species, including human hearts with and without characteristics of Brugada syndrome.
        %Note that the AP in the first human dataset does not have highly pronounced Brugada
        %syndrome features and instead is closer to a normal morphology. 
        Note that for the last case (human data with Brugada phenotype), only BOCF and BBOCF model fits are shown.
        All cases use \num{1024}
        particles and \num{32} iterations with no pre-recording stimuli except for the last case, which uses \num{16384} particles
        and \num{70} iterations, with the upper bounds for $\tau_{o1}$ and $\tau_{o2}$ increased to \num{1000}
        and \num{200}, respectively, for both models to accommodate the slow recovery time. The normalization constant was set to 1 for the BOCF model for the human data with Brugada phenotype, and to 1.05 and 1.35 for the BBOCF model for the normal and Brugada human cases, respectively.}
    \label{fig:multispecies}
\end{figure}

As further examples, Figure~\ref{fig:multispecies} shows fittings of the FK (blue), MS (orange), and BOCF (green) models to ventricular action potentials recorded from 
three different species: zebrafish, frog, and human, including a fairly typical AP and one showing prolongation with a shape characteristic of Brugada syndrome. While the frog and fish data were recorded
via microelectrode (paced at \qty{800}{\milli\second} and \qty{300}{\milli\second},
respectively), the human data was taken from an optical-mapping experiment performed on an
excised human heart (paced at \qty{1000}{\milli\second}). Despite these data showing a
wide variance in both their APDs and AP shapes, \ourtool found low-error parameterizations
for the FK, MS, and BOCF models, with the frog AP shape proving difficult for the FK model to match. In addition, the BBOCF model was also used to fit the
human tissue data, resulting in a better reproduction of the more complex morphology present in the human
dataset compared to the other models. For the Brugada phenotype, only the BBOCF model was capable of matching the ``saddleback'' shape; for comparison, Figure~\ref{fig:multispecies} also includes a fitting of this dataset obtained using the BOCF model.

Our fittings indicate key parameters involved in fitting action potentials from the four species (canine, fish, frog, and human) to the different models. In particular, we tested whether the parameters values obtained from 20 runs of \ourtool produced different parameter means and standard deviations for all six pairwise comparisons of the four species. For the MS model, $\tau_{in}$ (upstroke time scale) and $\tau_{close}$ (plateau time scale) were different for all pairwise comparisons; in fact, the only species pairs with parameter values that were not different were the frog and canine ($\tau_{open}$ and $v_{gate}$) and the fish and frog ($\tau_{out}$ and $v_{gate}$). 
For the FK model, there were no individual parameters whose values were different across all species comparisons, but parameters that were most commonly fit to different values across the species included $\tau_r$, $\tau_w^+$, $\tau_w^-$, $u_c$, and $u_c^{si}$.
For the BOCF model, the parameters most commonly involved in differentiating the action potentials for different species included $\theta_v$, $k_{so}$, $\tau_{s1}$, $\tau_{s2}$, $\tau_{w1}^+$, $\tau_{fi}$, and $u_u$. 
Details are given Tables~\ref{sup:tab:ms-comparison}-\ref{sup:tab:bocf-comparison}.

In addition, our results using the BBOCF model, which obtained high-quality fittings for both the normal and Brugada-phenotype action potentials, incdicated that several parameters play key roles in changing from the normal to Brugada-type shape. Specifically, the values obtained across 20 fittings for each case achieved different means for the following parameters ($p<0.001$): 
$\tau_{w2}^+$, $\tau_{s2}$, $\tau_{o2}$, $\tau_{si1}$, $\tau_{si2}$, $k_{so}$, $s_c$, and $u_{so}$. 
The parameter $k_{si}$ was also different for the two datasets at the level $p<0.01$. More description of the roles of these parameters can be found in Section~\ref{sec:discussion}.

\subsection{Fitting with APD data}

\begin{figure}[htpb]
    \centering
    \includegraphics{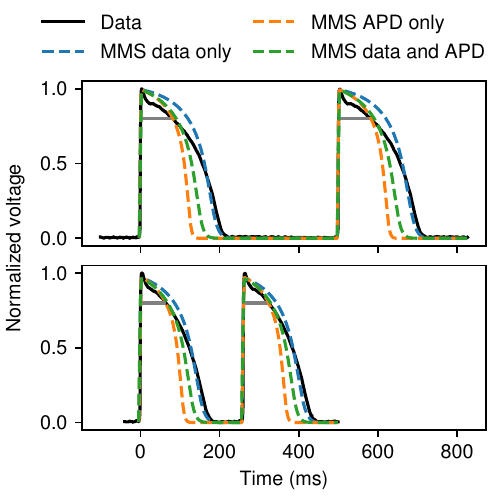}
    \caption{Fitting of the MMS model to canine endocardial data of cycle
    lengths \qty{260}{\milli\second} and \qty{500}{\milli\second}. The results of fitting
    to the voltage data only (blue), fitting to APD data only (orange), and a simultaneous fitting of voltage and APD data minimizing the APD error (green) are shown. The gray lines
    indicate the APD of the voltage data taken at a voltage value of \num{0.8}. All cases used \num{1024} particles and \num{32} iterations of PSO. Four pre-recording stimuli were used in all cases.}
    \label{fig:hybrid_fit}
\end{figure}

When matching the APD of a data set is particularly important, the APD fitting feature
of \ourtool can be used. This feature is particularly important for the two-variable models,
which otherwise do not emphasize APD accuracy and may achieve low curve error overall while having higher error during repolarization. 
An example of a fitting to both voltage and APD data is shown in Figure~\ref{fig:hybrid_fit},
where the MMS model was fit to canine endocardial data at CLs of \qty{500}{\milli\second} and \qty{260}{\milli\second}; APDs for this case were measured using a threshold of \num{0.8} corresponding to $\text{APD}_{20}$, which facilitates demonstration of the APD fitting feature. Fitting to voltage alone results in $\text{APD}_{20}$ errors of
\qty{46}{\percent} and \qty{35}{\percent} for the \qty{500}{\milli\second} and \qty{260}{\milli\second} cases respectively.
Fitting to the APD only resulted in respective APD errors of \qty{0}{\percent} and \qty{1}{\percent}, whereas
the hybrid fitting resulted in respective APD errors of \qty{10}{\percent} and \qty{8}{\percent}.
By simultaneously fitting the APD
%(in this case, $\text{APD}_{20}$, measured as the time above the voltage value of \num{0.8} for both data sets) 
and the voltage data, a
more accurate APD fitting was achieved at the cost of slightly higher curve error. The hybrid case used fitting weights of \num{1000} for
the APD data and \num{0.1} for the voltage trace.
\ourtool
also allows the option to fit APDs without using any voltage data, resulting in the highest
curve error, but giving more flexibility when limited data is available, such as for clinical data.

\subsection{Scaling and convergence}

\begin{figure}[htpb]
    \centering
    \includegraphics{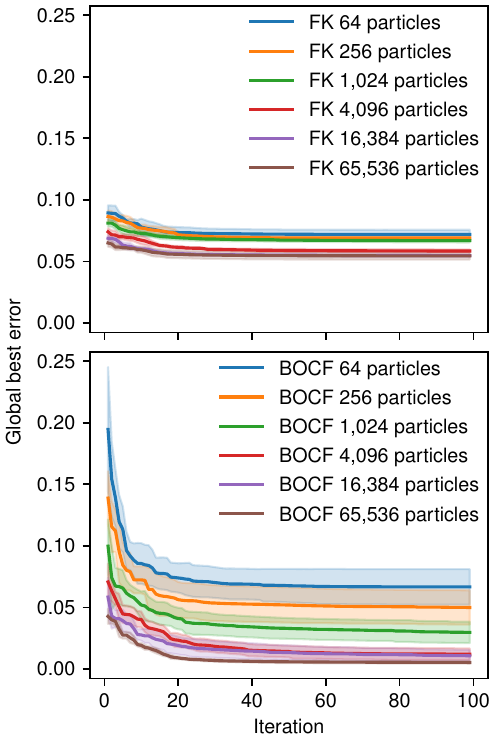}
    \caption{Global best error for each iteration averaged over five \ourtool runs. The shaded ribbons indicate
    the standard deviation of the error at that iteration over the five runs. Fits
    used the data from the four CLs of the Fox et. al model shown in Figure~\ref{fig:fox}. \ourtool
    algorithm was run for \num{1000} iterations, but only the first \num{100} are shown, as
    very little improvement occurred after this point.}
    \label{fig:err_fox}
\end{figure}

One of the advantages of using graphics hardware for parallelization is that \ourtool can
use a large number of particles for the PSO algorithm. Increasing the number
of particles generally led to a final result with lower error, although individual cases
did not always follow this trend due to the non-deterministic nature of the algorithm. 
We observed that increasing the number of particles is generally the most effective way to reduce
error when fitting a variety of datasets and models.
Examples of the error convergence behavior
of \ourtool using the Fox et al. model dataset fitted to the FK and BOCF models are shown in Figure~\ref{fig:err_fox}. 

Because the algorithm is guaranteed
to converge within a finite number of iterations (see Section~\ref{sec:pso-algorithm}), increasing
the number of iterations does not necessarily improve the result. However, the number of iterations
required for convergence varies depending on the model, number of particles, and data being
fit. When using \ourtool, the convergence plot shown at the bottom
of the page (see Figure~\ref{sup:fig:ui-error-plot}) can serve as a useful indicator of whether more
iterations would improve the fit.
In the cases shown in Figure~\ref{fig:err_fox}, increasing the number of iterations did not
compensate for using fewer particles. Although the error improved in a few cases after
many iterations, the improvement was slight relative to the improvement resulting from increasing
the number of particles. Figure~\ref{fig:err_fox} also demonstrates that the best error achieved (depicted by the shaded regions) for a given
choice of model, number of particles, and number of iterations could have significant variance, a
consequence of the non-determinism inherent in the PSO algorithm. 

\subsection{Program execution time}

One of the primary goals of \ourtool is to produce results quickly---ideally in a matter of
seconds---without sacrificing the quality of the fitting. Several factors affect the
running time of \ourtool beyond the implementation. Each iteration will take more time if more
simulation time is required to compare the model output to the data. Consequently, increasing the
number of data files, number of applied stimuli and pre-recording stimuli, number of cycle lengths, or complexity of the model to be fit
generally will increase
the running time. 
Additionally, increasing the number of iterations will
increase the running time, with each iteration taking a relatively consistent amount of time given
the other factors.
Increasing the number of particles also may increase the running time,
although the exact relationship between number of particles and time depends on the specific hardware
and drivers used to run \ourtool. 

\begin{figure}[htpb]
    \centering
    \includegraphics{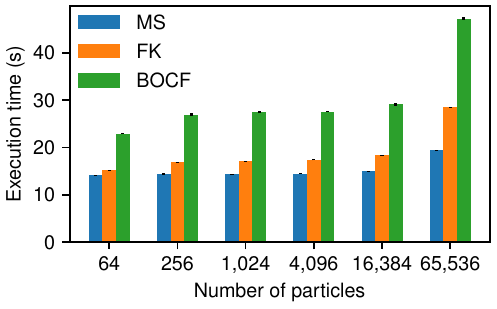}
    \caption{Average time over five runs of \ourtool to fit various models to the data
    shown in Figure~\ref{fig:fox} for varying numbers of particles. Each run used \num{100}
    iterations of the algorithm with four pre-recording stimuli and four stimuli applied. The standard deviation of the execution times is shown by the
    black line at the top of each bar.}
    \label{fig:timing}
\end{figure}

To demonstrate the speed of the program, as well as the effects of varying the number of particles, timing results are presented in Figure~\ref{fig:timing}
for fitting to the Fox et al.\ model data described in Section~\ref{sec:model-fit} with four stimuli and four pre-recording stimuli. Results were
recorded on a computer using an AMD Ryzen 9 7950X with 16 cores (32 virtual) with a clock speed of
\qty{5.881}{\giga\hertz}, an AMD RX 7900 XTX GPU, and \qty{64}{\giga\byte} RAM, running the Firefox
web browser. In this case,
the number of iterations was held constant at 100~iterations while the model and number of particles were varied. Even
the slowest case attempted, the BOCF model with \num{65536} particles, completed an average
run in less than \num{50} seconds. Model choice significantly impacts execution time; the BOCF model requires approximately double the time of the much simpler MS model
in many cases, with the FK model lying between the two. The number of particles also
plays a significant role, although the cost of a few seconds required to use \num{4096} particles as
opposed to \num{64} particles may be worth the higher-quality fits produced by more particles, especially when most modern graphics cards can comfortably accommodate thousands of particles.

The fact that the running time does not increase proportionately to the number of particles is
explained by the extreme parallelism offered by GPUs. In many cases, the GPU may have enough processing power that reducing the number of particles simply leaves available processing power
idle, which does not lead to speedup. Once the number of particles is large enough that the GPU
must work at full capacity, \ourtool will take longer to run as the number of particles is
increased further. However, the exact speeds and behavior will depend on many factors, most notably
the specific processor, graphics card, drivers, and browser version of the computer running the program.

\section{Discussion}
\label{sec:discussion}

\ourtool allows fast and easy fitting of a selection of cardiac models to both voltage time series
and APD data, which we have found to be successful for matching the output of both other models
and experimental data. We note in particular that users do not need to write or modify any code to
obtain fits from the tool, and all fits shown were obtained within a matter of seconds, even on laptops
without dedicated graphics hardware.
In addition, \ourtool does not require any machine-specific compilation or optimization, and may be run in a web
browser simply by loading the web page. Where possible, \ourtool provides end-users with sensible defaults for the
optimization hyperparameters and model constraints to minimize the effort required to produce a
model fit. Nevertheless, \ourtool users retain the
ability to control the hyperparameters of the PSO algorithm and constraints on the model through an
interactive user interface
Thus, \ourtool provides an experience that can meet the needs of both new and experienced users. In addition, the fittings that \ourtool obtains can provide insights into the roles of model parameters as well as the identifiability of the model given the provided data. 

\subsection{Insights into model parameter roles}

Our study helps to identify the parameters of individual models that support differentiating action potentials across species. For the MS model, which has a small number of parameters whose roles tend to be fairly specific and narrowly defined, most parameter values were found to be different across pairwise species comparisons (see Table~\ref{sup:tab:ms-comparison}). All parameter values obtained were different for the human dataset compared to the canine, fish, and frog data, and between the canine and fish datasets. Cases where values were not found to be different ($\tau_{open}$ and $v_{gate}$ for canine vs. frog and $\tau_{out}$ and $v_{gate}$ for fish vs. frog) could indicate either that the values were the same or that the variability was too high to make a meaningful distinction.

For the FK model, more variability in the parameters that were different was observed in pairwise species comparisons. The most commonly different parameter values were for $\tau_r$, which governs the strength of the repolarizing current; $\tau_w^+$ and $\tau_w^-$, which set the time constant of the slow inward current's gating variable and thus are primary contributors to action potential duration; and the threshold values $u_c$ and $u_c^{si}$. Interestingly, only three parameters were reliably distinguished between the canine and human datasets: $\tau_w^-$; $\tau_{v1}^-$, which affects how the fast inward current turns off; and $k$, which affects the slow inward current formation. Two parameters were found never to have meaningfully different values: $\tau_{v2}^-$ and $u_v$, which are both involved in the late stages of recovery of the $v$ gate and likely could not be pinpointed from the limited datasets.

The BOCF model similarly tended to find differences when comparing the fittings pairwise across species for variables affecting repolarizing current strength (e.g., $\tau_{s1}$, $\tau_{s2}$, and $k_{so}$), the fast inward current ($\tau_{fi}$, $\theta_v$, and $u_u$), and $\tau_{w1}^+$. Both the fish vs. canine and fish vs. frog comparisons had few meaningfully different values, whereas the human vs. canine comparison had statistically significantly different values for all but three parameters. As with the FK model, $\tau_{v2}^-$ was never found to be different across the pairwise species comparisons. 

Further insights can be obtained by comparing the BBOCF model parameter values obtained by fitting the normal and Brugada-type human action potentials. Of the nine parameters found to have statistically significantly different values for the two cases, four ($\tau_{si1}$, $\tau_{si2}$, $s_c$, and $k_{si}$) are used to define the time constant of the slow inward current (which is inversely proportional to the current's magnitude). The change in the magnitude of this current is consistent with the need to prolong the action potential for the Brugada phenotype. 
Parameters $k_{so}$ and $u_{so}$ affect the sigmoidal transition between the two time constants that affect the magnitude of the slow outward current for larger voltage values, thereby helping to form the saddleback feature. The remaining parameters affect the recovery of the activation and inactivation gating variables for the slow inward current, which also helps to prolong the action potential, as well as the magnitude of the slow outward current during the beginning of the late repolarization phase. Those parameters whose values were not statistically significant can be considered not important in producing the Brugada-phenotype action potential.

\subsection{Insights into model structure and parameter identifiability}

As can be seen in Figures~\ref{fig:self-recovery} and~\ref{fig:parameter-variance}, we found substantial variability in the values of a number of model parameters across repeated fittings to the same data, even though the voltage traces from the different parameterizations fit the data well
(see Figures~\ref{fig:self-3cl-ms}, \ref{fig:self-3cl-fk}, and~\ref{fig:self-3cl-bocf}). Lower variability typically was seen for some parameters of the MS model; for the FK and BOCF models, low variability generally was seen for some parameters, but the specific parameters with low variability depended on the dataset. Including more data through additional cycle lengths tended to decrease the variability in at least some parameters. 

The variability in parameter values is related to parameter identifiability, an area of ongoing research for cardiac models~\autocite{johnstone2016uncertainty,sher2022quantitative}, and may arise from a number of different model or data features. For example, there may not be sufficient data to constrain certain values, such as parameters associated with the minimum diastolic interval or conduction velocity properties (although the biphasic stimulus protocol helps to avoid at least some parameter combinations that would fail to propagate in tissue~\autocite{cairns2017}). Furthermore, a model may be limited by its structure and thus may not accommodate particular AP shapes, leading to irregular outcomes. In addition, the model may be insensitive to a particular parameter, or it may be possible for the model to compensate for an error in one parameter with an error in one or more other parameters within the context of the data to which the model is being fit. 
Fitting to different types of data, such as data generated from a stochastic pacing protocol~\autocite{krogh2016improving}, may be a valuable approach toward reducing variability in parameter values.

\subsection{Limitations and future work}
% NEED TO DISCUSS OTHER APPROACHES TO THE OPTIMIZATION PROBLEM (cite prior work) - e.g., genetic algorithms, combining global/local as in Loewe et al., etc.
A number of modifications or extensions could be made to \ourtool that could lead to improved performance or additional functionality. Currently, \ourtool offers the ability to fit data to six phenomenological models. However, allowing fittings to more detailed models, especially popular models of human ventricular cells~\autocite{ten2006alternans,ohara_simulation_2011}, would increase the \ourtool's usefulness. We intend to add these models to a future version of \ourtool, with the ability to fit maximal conductances of currents and possibly scale factors for gating variable time constants.

Our choice of PSO was motivated by certain advantages it provides: for example, adding a
model requires only its implementation as a simulation in WebGL shader code without a detailed analysis or transformation for the optimization problem, unlike 
linear regression or quadratic optimization approaches. 
%In contrast to past work using PSO or genetic algorithms, \ourtool is capable of fitting the entire parameter set of every model at once, without any assumptions about the solution required from end-users. 
In addition, PSO is capable of attaining rapid repeated fits on most
hardware due to its efficient parallelization of the PSO algorithm. 
Nevertheless, it would be possible to use a different fitting algorithm in addition to or instead of PSO. 
Although the speedup provided by
parallelization allows enough iterations and particles to be used in the PSO process that we
did not find it necessary to use further optimization methods on the result, as has been done
previously~\autocite{loewe2016parameter}, it would be possible to include a local optimization as a final step. Other approaches could be explored as well, including genetic algorithms~\autocite{bot_rapid_2012,cairns2017} and Bayesian approaches~\autocite{daly2015,nieto2023bayesian}.

As the behavior of the PSO algorithm
is closely coupled to the error metric used to evaluate the particles, one potential improvement
to our software would be to support different error metrics. For example, certain
parts of the voltage time series (e.g., the AP upstroke) could be weighted more heavily than others, thereby emphasizing specific lower-level features of the data beyond what the fitting weight parameters currently provided at the dataset level can support. Error metrics other
than squared error may also be beneficial; for instance, it has been suggested that the coefficient
of determination may be a more appropriate metric for regression problems~\autocite{chicco2021}.
The simple curve error currently utilized also may also lead to problems when data is not representative of
correct model behavior. For instance, optical-mapping data typically contain a slow upstroke due to
spatial averaging, which is not behavior models should attempt to reproduce. 
Currently, addressing upstroke blunting in optical-mapping data would require a separate tuning stage for upstroke parameters, whose values could then be held constant while other parameters were varied, but it is possible that greater improvement could be seen by excluding the upstroke from the error metric on the later fitting stages.

Additional potential improvements could be made to the data and protocols \ourtool supports. For example, an option to include a simultaneous calcium transient recording could be added, which could lead to improved fittings. 
The normalization constant could be included as a hyperparameter whose value is determined along with the model parameters as part of the optimization process.
Data obtained from protocols other than constant pacing at one or more fixed cycle lengths, such as an irregular pacing protocol~\autocite{groenendaal2015}, could be allowed and may present novel features that better constrain parameters.  Additional stimulus shapes could be incorporated beyond the square pulse and biphasic options. Another current limitation is that reaching short CLs without developing block may be difficult currently because the same initial conditions are used for each CL fit, but \ourtool could be modified to allow the final conditions of one CL to be used as initial conditions for the next shorter CL.
%In fact, arbitrary stimuli could be allowed by giving an option to upload a stimulus file similar to the data file. However, it is not clear if this feature would provide a significant benefit over the stimuli already available. An additional desired feature may be to allow an irregular pacing protocol for a single dataset, which is currently not supported in \ourtool. 
%Such a pacing protocol has been shown to improve parameter estimation in a genetic algorithm~\autocite{groenendaal2015}, but would require significant effort to implement and does not match any of the experimental data used in this study.

A significant future challenge is to consider tissue simulations, as models often behave differently
in simulated tissue when compared to single-cell simulations~\autocite{cherry2004suppression,clayton_models_2011}. Such a change would require a
significant increase in computation and in the complexity of \ourtool and would require careful consideration to avoid hindering \ourtool's speed and
simplicity from the end-user's perspective.
Finally, because \ourtool generates many sets of parameters that fit datasets relatively well, it could be extended in a different direction through modification to support the generation of populations of models~\autocite{britton2013experimentally} or parameter distributions~\autocite{daly2017,nieto2023bayesian}.

\section{Conclusion} 

We present \ourtool, a fast, user-friendly tool for identifying parameterizations of simple cardiac
models to closely match a set of input data. All model parameters, or any selected subset of parameters,
may be fit simultaneously, and a single parameterization may be generated to simultaneously fit multiple datasets taken
from different pacing rates. The speed and power of this software is made possible by the capabilities
of shader programs to take advantage of parallel graphics hardware available on almost all modern
consumer-grade computers. The fittings obtained can promote new insights into the roles of model parameters and the identifiability of models given specific data. Overall, we believe that \ourtool represents an important step toward making the
flexible and efficient phenomenological cardiac models practical and easy to use in patient-specific
applications for modeling experts and non-experts. 

\section*{Funding}

This study was supported by NSF grant CNS-2028677 and by NIH grants T32GM142616 and 2R01HL143450.

%Please refer to Journal-level guidance for any specific requirements.

%\section*{Funding}

%\begin{itemize}
%\item Funding: 
%This study was supported by NSF grants CNS-2028677 and CMMI-1762553 and by NIH grant 1R01HL143450.

%\item Conflict of interest/Competing interests: 
%(check journal-specific guidelines for which heading to use)

%The authors have no competing interests to declare that are relevant to the content of this article. The authors also declare they do not have any conflicts of interest.
%\item Ethics approval: 
%Not applicable.
%\item Consent to participate:
%Not applicable.
%\item Consent for publication:
%Not applicable.
%\item Availability of data and materials:
%The data used for this study are available upon request. 
%\item Code availability:
%Update:
%The code used for this study is available upon request. 
%\item Authors' contributions:
%\emph{This is optional and they give this example:}
%Update: All authors contributed to the study conception and design. Data collection and analysis were performed by ... The first draft of the manuscript was written by ... and all authors contributed to and commented on subsequent versions of the manuscript. All authors read and approved the final manuscript.
%\end{itemize}

\printbibliography
%\bibliography{pso-references}

\makeatletter\@input{xxx.tex}\makeatother

\end{document}

\title{Supplementary material}
\author{}
\date{}

\maketitle

\renewcommand{\thesection}{S\arabic{section}}
\renewcommand{\thefigure}{S\arabic{figure}}
\renewcommand{\thetable}{S\arabic{table}}
\renewcommand{\theequation}{S\arabic{equation}}

This supplement provides additional methods information, including the the detailed equations for each model as well as \ourtool's default bounds for all model parameters, a description of the available biphasic current, information on \ourtool advanced settings, and details on the parallel implementation of \ourtool. Furthermore, additional results in the form of significantly different parameters for pairwise species comparisons are included.

\section{Additional methods information}

\subsection{Model equations and parameter default bounds}
\label{sec:model-eqns}

\ourtool currently supports six models. The simplest is a modified version of the FitzHugh-Nagumo (MFHN)
model~\autocite{velasco2022methods}, which adds a second variable $v$ that evolves according to
\begin{align}
    \frac{dv}{dt} = \epsilon[(\beta-u) (u-\gamma) - \delta v - \theta],
\end{align}
and uses
\begin{align}
    I_\text{tot} = \mu u (1-u) (u-\alpha) - uv.
\end{align}
This model's seven parameters may be fit with \ourtool. Bounds for these parameters are given in Table~\ref{tab:mfhn-bounds}. The default value for the ``Normalize to'' field is 1.

\begin{table}[htpb]
    \centering
    \begin{tabular}{|c|c|c|}
        \hline
        Parameter & Minimum & Maximum\\
        \hline
        $\alpha$ & 0.05 & 0.6\\
        $\beta$ & 0.2 & 2\\
        $\epsilon$ & 0.001 & 1\\
        $\mu$ & 0.2 & 2\\
        $\gamma$ & 0.01 & 1\\
        $\theta$ & -0.1 & 0.1\\
        $\delta$ & 0.5 & 1.5\\
        \hline
    \end{tabular}
    \caption{Default parameter bounds for the MFHN model.}
    \label{tab:mfhn-bounds}
\end{table}

Another two-variable model available in \ourtool is the Mitchell-Schaeffer (MS) model~\autocite{mitchell2003}. This model uses
a gating variable $h$, which evolves according to the equation
\begin{align}
    \frac{dh}{dt} = \begin{cases}
        \frac{1-h}{\tau_\text{open}} & \text{if } u < v_\text{gate},\\
        \frac{-h}{\tau_\text{close}} & \text{if } u \geq v_\text{gate}.
    \end{cases}
\end{align}
The total current is separated into two components, $I_\text{tot}=I_\text{in}+I_\text{out}$, with
\begin{align}
    I_\text{in} &= \frac{h u^2 (1-u)}{\tau_\text{in}},\label{eq:ms-in}\\
    I_\text{out} &= -\frac{u}{\tau_\text{out}}.\label{eq:ms-out}
\end{align}
\ourtool's bounds for this model's five parameters are given in Table~\ref{tab:ms-bounds}. The default normalization constant is 1.

\begin{table}[htpb]
    \centering
    \begin{tabular}{|c|c|c|}
        \hline
        Parameter & Minimum & Maximum\\
        \hline
        $\tau_\text{in}$ & 0.15 & 0.6\\
        $\tau_\text{out}$ & 3 & 12\\
        $\tau_\text{close}$ & 75 & 300\\
        $\tau_\text{open}$ & 60 & 240\\
        $v_\text{gate}$ & 0.065 & 0.26\\
        \hline
    \end{tabular}
    \caption{Default parameter bounds for the MS model.}
    \label{tab:ms-bounds}
\end{table}

\ourtool also includes a modified version of the Mitchell-Schaeffer model (MMS)~\autocite{corrado2016two}, which modifies the
currents in Equations~\eqref{eq:ms-in} and~\eqref{eq:ms-out} to
\begin{align}
    I_\text{in} &= \frac{h u (u - v_\text{gate}) (1 - u)}{\tau_\text{in}},\\
    I_\text{out} &= -\frac{(1-h) u}{\tau_\text{out}}.
\end{align}
The MMS model retains the same five parameters as the MS model, with \ourtool's default bounds as given in Table~\ref{tab:mms-bounds} and the normalization constant set to 1.

\begin{table}[htpb]
    \centering
    \begin{tabular}{|c|c|c|}
        \hline
        Parameter & Minimum & Maximum\\
        \hline
        $\tau_\text{in}$ & 0.05 & 0.6\\
        $\tau_\text{out}$ & 0.5 & 12\\
        $\tau_\text{close}$ & 60 & 300\\
        $\tau_\text{open}$ & 60 & 240\\
        $v_\text{gate}$ & 0.065 & 0.26\\
        \hline
    \end{tabular}
    \caption{Default parameter bounds for the MMS model.}
    \label{tab:mms-bounds}
\end{table}

\ourtool also supports the three-variable Fenton-Karma (FK) model~\autocite{fenton1998}, which includes the variables $v$ and $w$:
\begin{align}
    \frac{dv}{dt} &= \begin{cases}
        \frac{1-v}{\tau_v^-(u)} & u < u_c,\\
        \frac{-v}{\tau_v^+} & u \geq u_c,
    \end{cases}\\
    \frac{dw}{dt} &= \begin{cases}
        \frac{1-w}{\tau_w^-} & u < u_c,\\
        \frac{w}{\tau_w^+} & u \geq u_c,
    \end{cases}
\end{align}
with
\begin{align}
    \tau_v^-(u) &= \begin{cases}
        \tau_{v2}^- & u < u_v,\\
        \tau_{v1}^- & u \geq u_v.
    \end{cases}
\end{align}
The total current is defined as 
$I_\text{tot}=I_\text{fi}+I_\text{so}+I_\text{si}$, where
\begin{align}
    I_\text{fi} &= \begin{cases}
        0 & u < u_c,\\
        -\frac{v}{\tau_d}(1-u)(u-u_c) & u \geq u_c,
    \end{cases}\\
    I_\text{so} &= \begin{cases}
        \frac{u}{\tau_o} & u < u_c,\\
        \frac{1}{\tau_r} & u \geq u_c,
    \end{cases}\\
    I_\text{si} &= -\frac{w}{2 \tau_\text{si}} (1 + \tanh[k(u-u_c^\text{si})]).
\end{align}
\ourtool allows fitting of all 13 model parameters, with default bounds specified in Table~\ref{tab:fk-bounds}. The normalization constant's default value for this model is 1.

\begin{table}[htpb]
    \centering
    \begin{tabular}{|c|c|c|}
        \hline
        Parameter & Minimum & Maximum\\
        \hline
        $\tau_r$ & 25 & 200\\
        $\tau_{si}$ & 10 & 300\\
        $\tau_w^+$ & 50 & 900\\
        $\tau_d$ & 0.15 & 0.4\\
        $\tau_v^+$ & 1 & 20\\
        $\tau_{v1}^-$ & 10 & 50\\
        $\tau_{v2}^-$ & 500 & 1500\\
        $\tau_w^-$ & 5 & 100\\
        $\tau_0$ & 5 & 50\\
        $k$ & 1 & 15\\
        $u_c^{si}$ & 0.2 & 0.9\\
        $u_c$ & 0.05 & 0.3\\
        $u_v$ & 0.005 & 0.06\\
        \hline
    \end{tabular}
    \caption{Default parameter bounds for the FK model.}
    \label{tab:fk-bounds}
\end{table}

The four-variable Bueno-Orovio-Cherry-Fenton
(BOCF) model~\autocite{bueno2008}, which supports more realistic AP shapes, is also implemented within \ourtool. It includes variables $v$, $w$, and $s$:
\begin{align}
    \frac{dv}{dt} &= \begin{cases}
        \frac{v_\infty(u)-v}{\tau_v^-(u)} & u < \theta_v,\\
        \frac{-v}{\tau_v^+} & u \geq \theta_v,
    \end{cases}\\
    \frac{dw}{dt} &= \begin{cases}
        \frac{w_\infty(u)-w}{\tau_w^-(u)} & u < \theta_w,\\
        \frac{-w}{\tau_w^+} & u \geq \theta_v,
    \end{cases}\\
    \frac{ds}{dt} &= \frac{(1+\tanh[k_s(u-u_s)])/2-s}{\tau_s(u)}.
\end{align}
Similar to the FK model, its total current is the sum of three currents: $I_\text{tot}=I_\text{fi}+I_\text{so}+I_\text{si}$, with
\begin{align}
    I_\text{fi} &= \begin{cases}
        0 & u < \theta_v,\\
        \frac{-v(u-\theta_v)(u_u-u)}{\tau_\text{fi}} & u \geq \theta_v,
    \end{cases}\\
    I_\text{so} &= \begin{cases}
        \frac{u-u_o}{\tau_o(u)} & u < \theta_w,\\
        \frac{1}{\tau_\text{so}(u)} & u \geq \theta_w,
    \end{cases}\\
    I_\text{si} &= \begin{cases}
        0 & u < \theta_w,\\
        \frac{-ws}{\tau_\text{si}} & u \geq \theta_w.
    \end{cases}
\end{align}
The other functions of $u$ in the BOCF equations are defined as follows:
\begin{align}
    \tau_v^-(u) &= \begin{cases}
        \tau_{v1}^- & u < \theta_v^-,\\
        \tau_{v2}^- & u \geq \theta_v^-,
    \end{cases}\\
    \tau_w^-(u) &= \tau_{w1}^- + (\tau_{w2}^--\tau_{w1}^-) \frac{1 + \tanh[k_w^-(u-u_w^-)]}{2},\\
    \tau_\text{so}(u) &= \tau_\text{so1} + (\tau_\text{so2}-\tau_\text{so1}) \frac{1 + \tanh[k_\text{so}(u-u_\text{so})]}{2},\\
    \tau_s(u) &= \begin{cases}
        \tau_{s1} & u < \theta_w,\\
        \tau_{s2} & u \geq \theta_w,
    \end{cases}\\
    \tau_o(u) &= \begin{cases}
        \tau_{o1} & u < \theta_o,\\
        \tau_{o2} & u \geq \theta_o,
    \end{cases}
\end{align}
and
\begin{align}
    v_\infty(u) &= \begin{cases}
        0 & u < \theta_v^-,\\
        1 & u \geq \theta_v^-,
    \end{cases}\\
    w_\infty(u) &= \begin{cases}
        \frac{1-u}{\tau_{w\infty}} & u < \theta_o,\\
        w_\infty^\ast & u \geq \theta_o.
    \end{cases}
\end{align}
The BOCF model uses 27 parameters in total, all of which are available to fit in \ourtool. Default parameter bounds for these parameters are shown in Table~\ref{tab:bocf-bounds}. The default value for the ``Normalize to'' field is \num{1.2}.

\begin{table}[htpb]
    \centering
    \begin{tabular}{|c|c|c|}
        \hline
        Parameter & Minimum & Maximum\\
        \hline
        $\theta_v$ & 0.1 & 0.35\\
        $\tau_{v1}^-$ & 0.5 & 300\\
        $\tau_{v2}^-$ & 1 & 1500\\
        $\tau_v^+$ & 1 & 15\\
        $u_w^-$ & 0.01 & 0.04\\
        $\tau_{so1}$ & 15 & 100\\
        $k_{so}$ & 1.8 & 2.2\\
        $\tau_{s1}$ & 2 & 3\\
        $\tau_{s2}$ & 1 & 20\\
        $k_s$ & 1.5 & 3\\
        $\tau_{w1}^-$ & 5 & 150\\
        $\tau_{w2}^-$ & 5 & 150\\
        $\tau_{w1}^+$ & 100 & 1000\\
        $\tau_{fi}$ & 0.05 & 0.5\\
        $\tau_{o1}$ & 5 & 500\\
        $\tau_{o2}$ & 5 & 10\\
        $\tau_{so2}$ & 0.1 & 1.5\\
        $u_{so}$ & 0.6 & 0.7\\
        $u_s$ & 0.8 & 1\\
        $\tau_{si1}$ & 1 & 4\\
        $\theta_w$ & 0.1 & 0.15\\
        $\theta_v^-$ & 0.005 & 0.25\\
        $\theta_o$ & 0.004 & 0.008\\
        $k_w^-$ & 50 & 250\\
        $\tau_{w_\infty}$ & 0.01 & 0.2\\
        $w_\infty^\ast$ & 0.4 & 1\\
        $u_u$ & 1.45 & 1.61\\
        \hline
    \end{tabular}
    \caption{Default bounds for the BOCF model.}
    \label{tab:bocf-bounds}
\end{table}

The final model included in \ourtool is a modification of the BOCF model for modeling action potentials in tissue
with Brugada syndrome (BBOCF)~\autocite{bueno2015basis}. This model modifies the BOCF model in two main ways. First, the Heaviside
functions in the equations for $v_\infty(u)$, $w_\infty(u)$, $I_\text{so}$, $I_\text{si}$, and $\tau_s$ are updated to
use independent threshold parameters $\theta_{v_\infty}$, $\theta_{w_\infty}$, $\theta_\text{so}$, $\theta_\text{si}$, and
$\theta_s$, respectively, in place of the parameters appearing in the other equations $\theta_v^-$, $\theta_o$,
$\theta_w$, $\theta_w$, and $\theta_w$. Second, two of the time constants are reformulated as functions of the gating
variables,
\begin{align}
    \tau_w^+(w) &= \tau_{w1}^+ + (\tau_{w2}^++\tau_{w1}^+) \frac{1+\tanh[k_w^+(w-w_c^+)]}{2},\\
    \tau_\text{si}(s) &= \tau_\text{si1} + (\tau_\text{si2}+\tau_\text{si1}) \frac{1+\tanh[k_\text{si}(s-s_c)]}{2}.
\end{align}
This modification results in a total of 39 parameters, all of which are available to fit in \ourtool using the default bounds given in Table~\ref{tab:bbocf-bounds}. The default value for the normalization constant is \num{1.2}.

% Some of the parameters here are fixed by setting the maximum and minimum to the same value, which we should probably either discuss or change.
\begin{table}[htpb]
    \centering
    \begin{tabular}{|c|c|c|}
        \hline
        Parameter & Minimum & Maximum\\
        \hline
        $\tau_{v1}^+$ & 5 & 10\\
        $\tau_{v1}^-$ & 60 & 60\\
        $\tau_{v2}^-$ & 50 & 50\\
        $\tau_{w1}^+$ & 40 & 80\\
        $\tau_{w2}^+$ & 150 & 300\\
        $\tau_{w1}^-$ & 10 & 500\\
        $\tau_{w2}^-$ & 20 & 40\\
        $\tau_{s1}$ & 5 & 15\\
        $\tau_{s2}$ & 50 & 90\\
        $\tau_{fi}$ & 0.05 & 0.1\\
        $\tau_{o1}$ & 400 & 500\\
        $\tau_{o2}$ & 20 & 35\\
        $\tau_{so1}$ & 150 & 200\\
        $\tau_{so2}$ & 1 & 3\\
        $\tau_{si1}$ & 10 & 20\\
        $\tau_{si2}$ & 2 & 10\\
        $\tau_{w_\infty}$ & 0.12 & 0.12\\
        $\theta_v$ & 0.13 & 0.13\\
        $\theta_v^-$ & 0.006 & 0.006\\
        $\theta_{v_\infty}$ & 2 & 2\\
        $\theta_w$ & 0.13 & 0.13\\
        $\theta_{w_\infty}$ & 0.12 & 0.12\\
        $\theta_{so}$ & 0.2 & 0.2\\
        $\theta_{si}$ & 0.13 & 0.13\\
        $\theta_o$ & 0.006 & 0.006\\
        $\theta_s$ & 0.36 & 0.36\\
        $k_w^+$ & 5 & 10\\
        $k_w^-$ & 100 & 150\\
        $k_s$ & 5 & 25\\
        $k_{so}$ & 1.5 & 4\\
        $k_{si}$ & 10 & 70\\
        $u_w^-$ & 0.02 & 0.12\\
        $u_s$ & 0.25 & 0.4\\
        $u_o$ & 0 & 0\\
        $u_u$ & 1 & 1\\
        $u_{so}$ & 0.3 & 0.75\\
        $s_c$ & 0.6 & 0.9\\
        $w_c^+$ & 0.2 & 0.3\\
        $w_\infty^\ast$ & 0.94 & 0.94\\
        \hline
    \end{tabular}
    \caption{Default parameter bounds for the BBOCF model. Note that these bounds keep 15 parameters constant by setting the upper and lower bounds to be equal; the bounds can be changed in the interface to include those parameters in the fitting.}
    \label{tab:bbocf-bounds}
\end{table}

\subsection{Biphasic stimulus}
\label{sec:biphasic-stim}

\begin{figure}[htpb]
    \centering
    \includegraphics{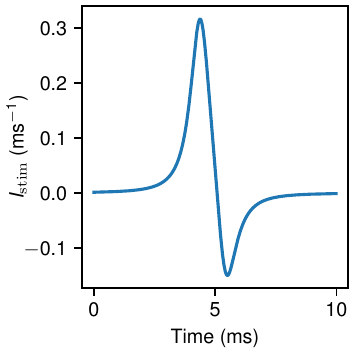}
    \caption{Biphasic stimulus generated from default values that is available for pacing as an alternative to a square pulse.}
    \label{fig:stim}
\end{figure}

The biphasic stimulus has the form
\begin{align}
    I_\text{stim} = -I_\text{mag} \times
    \frac{\frac{t}{a} - b}{1 + \left(\frac{t}{a} - c\right)^4},
    \label{eq:biphasic-stimulus}
\end{align}
%\begin{align}
%    I_\text{stim} = -I_\text{mag} \times
%    \frac{\frac{t}{t_\text{scale}} - t_\text{offset1}}{1 + \left(\frac{t}{t_\text{scale}} - t_\text{offset2}\right)^4},
%    \label{eq:biphasic-stimulus}
%\end{align}
where $t$ is the time in milliseconds since the stimulus began. The default parameter values for the biphasic
stimulus are $I_\text{mag}=\qty{0.4}{\per\milli\second}$, $a=\qty{0.725}{\milli\second}$,
$b=7$, and $c=6.72$, applied for a duration of \qty{10}{ms}. This stimulus current with the default parameter values is
shown in Figure~\ref{fig:stim}. All parameters controlling the two types of stimuli can be modified in the user interface,
as described in Section~\ref{sec:interface-stimulus}.

\subsection{\ourtool advanced settings}

\subsubsection{Specifying the stimulus type and parameters}
\label{sec:interface-stimulus}

The stimulus may be chosen as either a square or biphasic stimulus.
The stimulus settings are located below the section for adding cardiac data. If a square stimulus is chosen
(the default), only ``Stimulus duration'' and ``Stimulus magnitude'' appear as options to be changed if
desired. To switch to (from) the biphasic stimulus, check (uncheck) the ``Biphasic stimulus'' checkbox. If the biphasic stimulus is selected,
additional options will appear corresponding to the parameters in Equation~\eqref{eq:biphasic-stimulus} in the Supplement
The interface for setting the parameters of the square stimulus are shown in Figure~\ref{main:fig:squarestim-ui}, and the
interface for setting the parameters of the biphasic stimulus are shown in Figure~\ref{main:fig:biphasicstim-ui}. In the
biphasic stimulus equations given in Section~\ref{sec:biphasic-stim}, $b$ is controlled by ``Stimulus offset 1'', $c$
is controlled by ``Stimulus offset 2'', and $a$ is controlled by ``Stimulus timescale''.

\subsubsection{Setting PSO hyperparameters and monitoring convergence}

%\begin{figure}[htpb]
%    \centering
%    \includegraphics[width=84mm]{Figures/ui_plot.png}
%    \caption{\ourtool user interface plot of the fitted model. The user-provided dataset is shown in black and the model
%        fit is shown in red.}
%    \label{fig:ui-plot}
%\end{figure}

\begin{figure}[htpb]
    \centering
    \includegraphics[width=50mm]{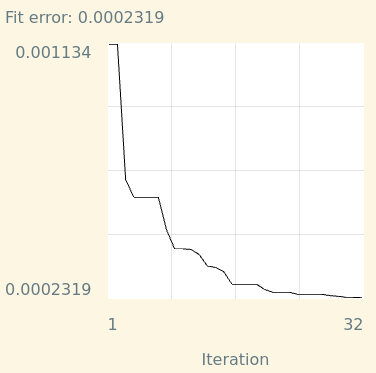}
    \caption{\ourtool user interface plot of the lowest error found (vertical axis) up to each iteration (horizontal axis). The
        error corresponding to the best fit is shown above.}
    \label{fig:ui-error-plot}
\end{figure}

Settings to change the default \ourtool hyperparameters described in Section~\ref{main:sec:pso-algorithm} 
are located
at the bottom of the page, shown in Figure~\ref{main:fig:ui-hyperparams}. The local uniform maximum corresponds to $\phi_1$
in the equations, the global uniform maximum corresponds to $\phi_2$, and the constriction coefficient corresponds to
$\chi$. Increasing the number of particles and/or the number of iterations may lead to improved fit quality but is not
guaranteed to do so, due to the randomness in the algorithm. However, using more particles than the graphics hardware
can support will result in longer fitting times.

It is often useful to run \ourtool a few times to test the sensitivity of the results to particle initialization; clicking the
``Run'' button again will do so without the need to make any adjustments on the interface.
The plot of the lowest error found, as depicted in Figure~\ref{fig:ui-error-plot}, can be used to monitor convergence of the algorithm.

\subsection{\ourtool parallel implementation details}
\label{sec:parallel-impl}

\ourtool exploits the large-scale parallelism provided by graphics
hardware by using the WebGL API for JavaScript. This API is used to run parallel programs and automatically
leverages any hardware acceleration present on the host machine without requiring compilation or
manual tuning by the user of the software. WebGL is natively supported by all modern web browsers;
all that is required to run the program is to visit a website (either running on the user's computer if the code has been
downloaded, or over the Internet).

The WebGL API uses two-dimensional image
textures as its primary data structure, with a specific set of texture coordinates referred to as a ``texel.''
Each particle is mapped by its index to a unique location on the texture grid. 
For the update step, the input texels contain the current positions of each
particle, and the output texels contain the computed fit error of each particle.

When using \ourtool to fit multiple data sets (e.g., voltage time series obtained from pacing at different cycle lengths), the weighted sum of each fit error is aggregated using the blend function
of the WebGL API, a built-in feature of WebGL that allows element-wise operations on entire textures to be
performed efficiently without a custom shader implementation. Position and velocity update steps are performed in parallel as well, as this update step
is fully independent from other particles once global values such as $\vec{b}$ have been computed.

When computing global values in \ourtool, parallel speedup is achieved by splitting
the operation into two steps. First, the lowest error along one axis is found, for example, by comparing texels
with matching $x$-indices. The second step compares the results along the other axis, so that every texel
now has the best value and its corresponding index. A benefit of this approach is that these computations can
be performed on the GPU, and the resulting values are immediately available as texture data for the particle and
velocity update step. Consequently, the computation for each iteration requires little speedup-limiting communication
between the GPU and CPU.

\section{Additional results}

Tables~\ref{tab:ms-comparison}-\ref{tab:bocf-comparison} indicate parameter values that are different at the $p<0.01$ and $p<0.001$ levels of significance for pairwise species comparisons using the MS, FK, and BOCF models. In all cases, 20 fittings were compared for the single-CL fits of the fish, frog, and human (normal) data and the four-CL fit for the canine data.

\begin{table}[htpb]
    \centering
    \begin{tabular}{|c|c|c|c|c|c|c|}
        \hline
        Parameter & C vs. Fi & C vs. Fr & C vs. H & Fi vs. Fr & Fi vs. H & Fr vs. H\\
        \hline
        $\tau_\text{in}$    & ** & ** & ** & ** & ** & ** \\
        $\tau_\text{out}$   & ** & ** & ** &    & ** & ** \\
        $\tau_\text{close}$ & ** & ** & ** & ** & ** & ** \\
        $\tau_\text{open}$  & ** &    & ** & ** & ** & ** \\
        $v_\text{gate}$     & ** &    & ** &    & ** & ** \\
        \hline
    \end{tabular}
    \caption{Parameter values for pairwise species comparisons for the MS model that are different with $p<0.001$~(**). C~=~Canine, Fi~=~Fish, Fr~=~Frog, H~=~Human (normal).}
    \label{tab:ms-comparison}
\end{table}

\begin{table}[htpb]
    \centering
    \begin{tabular}{|c|c|c|c|c|c|c|}
        \hline
        Parameter & C vs. Fi & C vs. Fr & C vs. H & Fi vs. Fr & Fi vs. H & Fr vs. H\\
        \hline
        $\tau_r$      & ** & ** &    & ** & ** & ** \\
        $\tau_{si}$   & *  & ** &    & ** &    & ** \\
        $\tau_w^+$    & ** & ** &    &    & ** & ** \\
        $\tau_d$      & ** & ** &    & *  & ** &    \\
        $\tau_v^+$    & ** & *  &    &    & ** & ** \\
        $\tau_{v1}^-$ &    &    & ** &    & ** & *  \\
        $\tau_{v2}^-$ &    &    &    &    &    &    \\
        $\tau_w^-$    &    & ** & ** &    & ** & ** \\
        $\tau_0$      &    & ** &    & ** &    & ** \\
        $k$           & ** &    & ** & *  &    & *  \\
        $u_c^{si}$    & ** &    &    & ** & ** & ** \\
        $u_c$         & ** & ** &    & ** & ** & ** \\
        $u_v$         &    &    &    &    &    &    \\
        \hline
    \end{tabular}
    \caption{Parameter values for pairwise species comparisons for the FK model that are different with $p<0.01$~(*) or $p<0.001$~(**). C~=~Canine, Fi~=~Fish, Fr~=~Frog, H~=~Human (normal).}
    \label{tab:fk-comparison}
\end{table}

\begin{table}[htpb]
    \centering
    \begin{tabular}{|c|c|c|c|c|c|c|}
        \hline
        Parameter & C vs. Fi & C vs. Fr & C vs. H & Fi vs. Fr & Fi vs. H & Fr vs. H\\
        \hline
        $\theta_v$        &    & ** &    & ** & ** & ** \\
        $\tau_{v1}^-$     &    & *  & ** &    & ** & *  \\
        $\tau_{v2}^-$     &    &    &    &    &    &    \\
        $\tau_v^+$        & *  & ** & ** &    &    & *  \\
        $u_w^-$           &    &    & ** &    & *  & ** \\
        $\tau_{so1}$      &    &    & ** &    &    & *  \\
        $k_{so}$          &    & ** & ** &    & ** & ** \\
        $\tau_{s1}$       &    & ** & ** &    & ** & ** \\
        $\tau_{s2}$       & ** & ** & ** & ** &    &    \\
        $k_s$             & ** &    & ** &    &    & ** \\
        $\tau_{w1}^-$     &    &    & ** &    &    &    \\
        $\tau_{w2}^-$     &    & *  & ** &    &    &    \\
        $\tau_{w1}^+$     & ** & ** &    & ** & *  & ** \\
        $\tau_{fi}$       &    & ** & ** &    & ** & ** \\
        $\tau_{o1}$       &    &    & ** &    &    &    \\
        $\tau_{o2}$       &    & ** & ** &    & ** &    \\
        $\tau_{so2}$      & *  &    & ** &    &    &    \\
        $u_{so}$          &    & ** & ** & *  &    & ** \\
        $u_s$             &    &    & ** &    &    &    \\
        $\tau_{si1}$      &    & ** & ** &    & ** & *  \\
        $\theta_w$        &    & *  & ** &    & ** & ** \\
        $\theta_v^-$      &    &    & ** &    & *  & *  \\
        $\theta_o$        &    & *  & ** &    &    & *  \\
        $k_w^-$           &    & ** & ** &    & ** & *  \\
        $\tau_{w_\infty}$ &    &    & ** &    &    &    \\
        $w_\infty^\ast$   &    & ** & ** &    & ** &    \\
        $u_u$             &    & ** & ** &    & ** & ** \\
        \hline
    \end{tabular}
    \caption{Parameter values for pairwise species comparisons for the BOCF model that are different with $p<0.01$ (*) or $p<0.001$ (**). C~=~Canine, Fi~=~Fish, Fr~=~Frog, H~=~Human (normal).}
    \label{tab:bocf-comparison}
\end{table}

\printbibliography

\makeatletter\@input{xx.tex}\makeatother